# A multi-stop time-of-flight spectrometer for the measurement of positron annihilation-induced electrons in coincidence with the Doppler-shifted annihilation gamma photon

V. A. Chirayath[1]*, R. W. Gladen[1]┼, A. D. McDonald[1], A. J. Fairchild[1], P. V. Joglekar[1], S. Satyal[1], Z. H. Lim[1], T. N. Shead[1], M. D. Chrysler[1], S. Mukherjee[1,2], B. M. Barnett[1], N. K. Byrnes[1], A. R. Koymen[1], R. G. Greaves[3], and A. H. Weiss[1]

[1]*Department of Physics, University of Texas at Arlington, Arlington, Texas 76019, USA.*
[2]*Radiochemistry Division, Bhabha Atomic Research Center, Trombay, Mumbai 400085, India.*
[3]*First point Scientific Inc., Agoura Hills, California 91301, USA.*

*) Corresponding author: chirayat@uta.edu, ┼) Corresponding author: randall.gladen@uta.edu

**Abstract.** Here we describe an advanced multi-functional, variable-energy positron beam system capable of measuring the energies of multiple 'positron-induced' electrons in coincidence with the Doppler-shifted gamma photon resulting from the annihilation of the correlated positron. The measurements were carried out using the unique characteristics of the digital time-of-flight spectrometer and the gamma spectrometer available with the advanced positron beam system. These measurements have resulted in (i) the first digital time-of-flight spectrum of positron annihilation-induced Auger electrons generated using coincident signals from a high-purity Ge detector and a micro-channel plate; (ii) a two-dimensional array of the energy of Doppler-broadened annihilation gamma and the time-of-flight of positron-annihilation induced Auger electrons/secondary electrons measured in coincidence with the annihilation gamma photon; and (iii) the time-of-flight spectra of multiple secondary electrons ejected from a bilayer graphene surface as a result of the impact and/or annihilation of positrons. The novelty of the gamma – electron coincidence spectroscopy has been demonstrated by extracting the Doppler-broadened spectrum of gamma photons emitted due to the annihilation of positrons exclusively with 1s electrons of carbon. The width of the extracted Doppler-broadened gamma spectrum has been found to be consistent with the expected broadening of the annihilation gamma spectrum due to the momentum of the 1s electrons in carbon.

## 1. INTRODUCTION

When a thermalized positron annihilates with an electron, the momentum, **p**, of the annihilating electron-positron pair imparts a small deviation to the collinearity of the emitted photons and produces a measurable shift (red shift/blue shift) to the energy of the annihilation gamma photon as a result of the Doppler effect [1-5]. Since the total momentum of the annihilating electron-positron pair is dominated by the momentum of the electron, a measurement of the angular correlation or the Doppler broadening of annihilation radiation can be used to determine the momentum distribution of electrons at the site of positron annihilation. The measured Doppler-broadened photo-peak contains within it broadening due to the momentum distribution of valence electrons as well as the core electronic levels in the sample. These partial contributions to the total Doppler broadening are not usually resolved experimentally but are, instead, estimated theoretically using a calculation of the Doppler broadening to the annihilation gamma due to the momentum distribution of each electronic state. The calculated annihilation gamma spectrum corresponding to core electronic level momentum distributions are scaled by their respective calculated annihilation rates and added together to obtain the full Doppler-broadened gamma spectrum from the sample [6]. The contributions to the Doppler broadening from valence electron annihilations are typically not added to the theoretical spectrum due to the lack of an accurate description of the valence electron states. Since the Doppler broadening of the annihilation gamma depends critically on the local electronic structure, elemental composition, and defect architecture, it is extremely difficult to check the validity of the calculated partial contributions to the total Doppler-broadened spectrum [7]. However, it is possible to extract the partial contributions experimentally by measuring the annihilation gamma photons in coincidence with the Auger electrons or x-rays that are emitted following the decay of the positron annihilation-induced core holes.

Eshed et al. [8] first demonstrated a method in which the energy of the gamma photon produced by electron-positron annihilation was measured in coincidence with the electron emitted by the Auger decay of an annihilation induced core hole. Specifically, the annihilation-induced gamma photons were measured in coincidence with Auger electrons emitted following the decay of 3p (4p) annihilation-induced holes in Cu (Ag). This technique resulted in the first direct measurement of the momentum distribution of 3p (4p) electrons in Cu (Ag) and the results of the experiment were compared to theoretical calculations. The measurements however were performed with the use of a trochoidal electron energy analyzer; consequently, the Doppler-shifted annihilation gammas were measured in coincidence with Auger electrons emitted within a narrow range of energies. As a result of this restriction, Eshed et al. [8] and, later, Kim et al. [9] measured momentum distributions pertaining only to one element and one atomic orbital, limiting the applicability of the gamma-Auger electron coincidence measurements in situations where there are multiple elements and thus multiple Auger peaks. We describe here an advanced positron beam that has been developed with the primary objective of measuring the energies of the Doppler-shifted gamma photons in coincidence with the electrons that are emitted following positron impact and/or annihilation. Our present apparatus is equipped with a magnetic-field parallelizer resulting in a $2\pi$ electron detector [10] attached to a 3 m time-of-flight (ToF) spectrometer, and a high-resolution gamma spectrometer, allowing the measurement of the ToF of positron-induced electrons emitted with energies from 0 eV to 1000 eV and the energy of the Doppler-shifted gamma emitted following the annihilation of a positron associated with the detected electron-emission event.

In a ToF spectrometer, the energy of a positron-induced electron is obtained from the time difference between the detection of the electron at a micro-channel plate (MCP) detector and the detection of the annihilation gamma by a gamma detector. The ToF spectrometer can simultaneously collect electrons emitted from the surface into $2\pi$ sr and within a wide range of energies, making the coincident measurement of electron energies and annihilation gamma energies highly efficient. The 3 m flight path in the present apparatus permits the timing information of the gamma to be derived from the signals produced by the same high-resolution gamma detector that measures the energy distribution of the annihilation gamma. Deriving the time and energy information pertaining to the annihilation gamma from the same detector signal enables the construction of a two-dimensional array of correlated positron-induced electron ToFs and annihilation gamma energies. The multi-faceted information contained within this two-dimensional spectrum can be derived by taking appropriate cuts along the time or energy axes. We demonstrate here the novelty of the gamma-electron coincidence technique developed by deriving the gamma spectrum originating exclusively from the annihilation of positrons with the 1s electrons of carbon.

The digitization of the acquisition of the electron and gamma-induced signals in the respective detectors additionally allowed us to measure the ToF spectra of multiple secondary electrons that were emitted following the implantation of energetic positrons; and now the single-stop analog ToF spectrometer has been converted into a multi-stop digital ToF spectrometer that measures all electron emissions associated with a single positron implantation/annihilation event. We will discuss an application of this multi-stop ToF spectrometer and the resulting ToF spectra of multiple secondary electrons.

The advanced positron beam is an upgraded version of a previous ToF-PAES system at the University of Texas at Arlington [11]. The main advancements are in the source-moderator assembly, the digitization of the data acquisition methodology, the extended flight path and the non-conventional use of a high energy resolution Ge detector for the determination of the ToF of electrons. In the following sections, we highlight how the modifications resulted in the ability to coincidently measure the energies of 'positron-induced' electrons and the energies of the correlated annihilation gamma. The coincidence measurement capabilities, when applied to PAES, are expected to be useful in the study of emergent properties on the surfaces of topological insulators [12] or graphene [13] and in the study of new processes involving positron-surface interactions [14-18]. The application of the multi-electron ToF spectrometer to measure the energies of positron annihilation-induced Auger electrons is expected to resolve the contribution of multi-electron Auger decay pathways of the core or deep valence holes which in turn can give critical information on the strength of electron correlation [19-24].

In the following section we give a detailed description of the experimental system including the positron beam energy, the timing characteristics of the 3 m ToF spectrometer, and the energy resolution. In section 3 we describe the first results obtained with the novel capabilities of the advanced positron beam including the two-dimensional gamma electron coincidence spectrum and the spectra of multiple secondary electrons followed by possible applications of the multifunctional positron beam.

## 2. DESCRIPTION OF THE APPARATUS

The drawing of the advanced positron beam system with a 3m flight path is shown in Figure 1. Figure 1(a) shows the cross-sectional view of the advanced positron beam system except for the support structure and the transverse coils used for the correction of stray magnetic fields. Figure 1(b) and Figure 1(c) shows the sample chamber and the region near the electron detection system, respectively, in an expanded scale. The main components of the advanced positron beam system are (i) the positron source and moderator assembly, (ii) the $\bar{E} \times \bar{B}$ system, (iii) the electron detector, (iv) the field-free ToF tube, and (v) the sample chamber. We will discuss the details of the apparatus in the following sub-sections.

### 2.1. *Positron source and moderator assembly*

For the experiments reported here, Na-22 (~ 6.9 MBq) produced the positrons through positive beta decay. The advanced positron beam system utilizes a rare gas moderator (RGM) system designed and manufactured by First point Scientific Inc. [25] to obtain a mono-energetic beam of slow positrons. Specifically, the RGM utilizes solidified Neon to obtain moderated mono-energetic slow positrons from the fast-positive beta decay spectrum. The Na-22 source capsule is mounted in a copper holder with conical face, which in turn is attached to a closed cycle refrigerator. Using this refrigerator, the temperature of the copper holder can be reduced to as low as 5 K. The moderator is grown on the conical face by spraying Ne gas onto the Cu holder which is kept at 8.8 K during the moderator growth. The solid Ne layer, thus grown on the Cu holder, is further annealed at 9.3 K to remove the positron-trapping open volume defects from the thin film. During normal beam operation, the temperature of the Cu holder is maintained at 6.8 K. For the experiments reported here, new moderators were grown every two days, but the moderator has been tested to be efficient for approximately seven days with an RGM base pressure of $2.7 \times 10^{-7}$ Pa. Upon reemission from the Ne surface, the slow positrons are separated from the fast, non-moderated positrons by bending the slow positrons through a tungsten barrier with an offset aperture as shown in Figure 2. The bending is achieved by adding a small transverse field component to the axial magnetic field used for transport. The energy-filtered slow positrons pass through a pumping restriction, which has a diameter of 6.35 mm, before entering the subsequent $\bar{E} \times \bar{B}$ system. The initial positron beam diameter is thus restricted by the pumping restriction. We operate the RGM system at two different settings, as given in Table 1, to obtain slow positrons of desired energy at the sample. The efficiency of the RGM attached to the advanced positron beam, defined as number of slow positrons per $\beta^+$ emission from the source, measured at the entrance of the $\bar{E} \times \bar{B}$ system is ~ 0.2% [25, 26]. With a 6.9 MBq source and 0.2% moderation efficiency, the advanced positron beam takes approximately 30 days to collect gamma-Auger coincidence spectrum with reasonable statistics. With a higher source activity, we expect a proportional reduction in the data collection time.

|  | *Low energy settings* | *High energy settings* |
|---|---:|---:|
| Helmholtz Coil | 9.5 mT | 12.5 mT |
| Solenoid | 25.1 mT | 30.0 mT |
| Saddle Coils (1, 1', 2, 2') | 2.8 mT | 3.2 mT |
| Moderator bias | +3 V | +15 V |

**Table 1**: Settings used at the RGM system to obtain monoenergetic positrons of required energy.

|  | Low energy settings | | | High energy settings | |
| --- | --- | --- | --- | --- | --- |
|  | North | South |  | North | South |
| $\bar{E} \times \bar{B}$ set 1 (East) | 0 | -85 V | $\bar{E} \times \bar{B}$ set 1 (East) | 0 | -85 V |
| $\bar{E} \times \bar{B}$ set 2 (West) | -11.3 V | +10.3 V | $\bar{E} \times \bar{B}$ set 2 (West) | -25.1 V | +27.3 V |
| ToF tube bias | -0.5 V / 0 V | | ToF tube bias | -68 V / 0 V | |
| Sample bias | -1 V to -20 kV | | Sample bias | -25 to -20 kV | |

**Table 2**: Settings used for the transport of positrons to the sample and the transport of electrons to the MCP. The ToF tube bias and the sample bias are modified depending on the experiment to be performed.

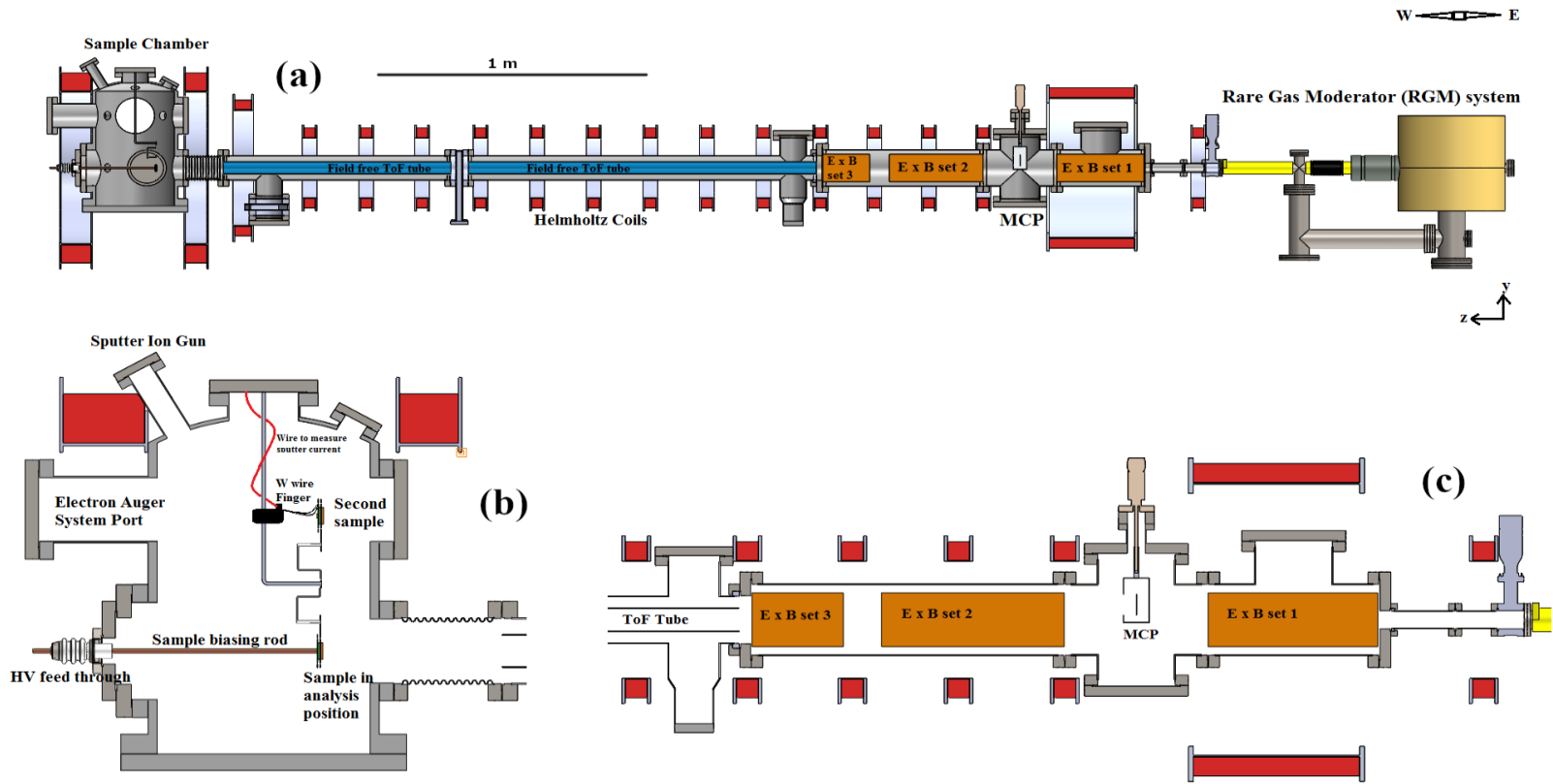

**Figure 1**. (a) Cross sectional view of the advanced positron beam system. The support structure and the Helmholtz coils used to produce the transverse magnetic fields are not shown. The main components of the beam system are the RGM system that houses the source, the $\bar{E} \times \bar{B}$ system, the Helmholtz coils, the ToF tube and the sample chamber. (b) Detailed view of the sample chamber showing the sample holder with two samples. The ports for Sputter ion gun and the electron-induced Auger electron system are marked. The biasing rod and the wire connection to measure the current during the sputtering is also shown. (c) Detailed view of the $\bar{E} \times \bar{B}$ system which consists of three pairs of plates that produce the electric field in the x direction. The Helmholtz coils produce the field in the z direction. The potential applied is such that the positron undergoes a drift along the negative y direction during its flight through the first set of plates and undergoes a drift along the positive y direction during its flight through the second set of plates. The third set of plates were grounded during the present measurements. However, it can be used to increase the path length of the charged particle through the $\bar{E} \times \bar{B}$ system if required.

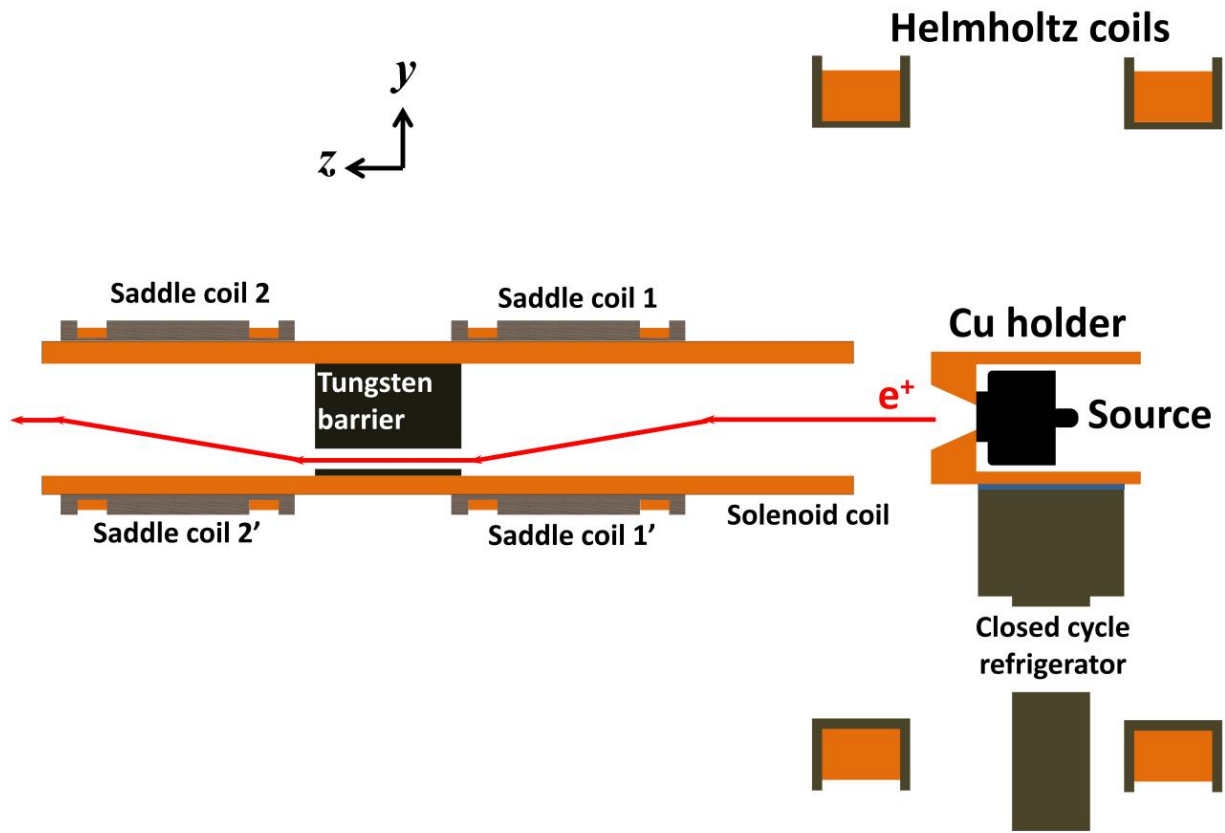

**Figure 2**. Schematic of the magnetic energy filtering in the RGM. A solid film of Ne is grown on the conical face of the Cu holder that is maintained at a few Kelvin. The slow mono-energetic positrons emitted from the Ne thin film are separated from the fast-non-moderated positrons using magnetic energy filtering in RGM. The slow positrons are transported along the beam axis using the magnetic field generated along the z axis by a combination of solenoid and Helmholtz coils. Four coils saddled on the solenoid coil add a transverse field component to the axial magnetic field. Saddle coils 1 and 1' create a field along the negative y direction whereas Saddle coils 2 and 2' create a field along the positive y direction. The magnetic field intensity of transverse field component is roughly a tenth of the intensity of the axial magnetic field. Only low energy positrons can make the bend and pass through the aperture whereas the fast positrons annihilate on the tungsten barrier.

## 2.2. Positron transport and the $\bar{E} \times \bar{B}$ system

The slow positrons from the RGM are magnetically transported to the sample using axial fields generated using a series of Helmholtz coils whose position along the beam line has been shown in Figure 1. The calculated axial magnetic field from the source to the sample is shown in Figure 3. The axial magnetic field is maintained at ~ $5 \times 10^{-3}$ tesla through most of the transport path. The axial field near the sample surface is, however, maintained at ~ $7.5 \times 10^{-2}$ tesla using a permanent magnet (SmCo alloy) that is placed behind the sample. This high magnetic field at the sample leads to the adiabatic compression of the positron beam [10, 27], resulting in a smaller beam spot (~ 2 mm diameter) on the sample. The positron transport is also aided by small transverse magnetic fields (~ $1 \times 10^{-4}$ tesla) applied throughout the length of the system to nullify any stray magnetic fields, including the magnetic field of Earth, that can deflect the low energy positrons from its axial path.

The slow positrons from the RGM that enter the main beam line are bent around the electron detector using the $\bar{E} \times \bar{B}$ system as shown in Figure 1 (c) and Figure 4. The $\bar{E} \times \bar{B}$ system consists of rectangular plates that can be biased to produce horizontal electric fields that are perpendicular to the axial magnetic field. The $\bar{E} \times \bar{B}$ field is thus, obtained by combining the axial magnetic field to the horizontal transverse electric field generated by two pairs of rectangular plates placed before (on the east side) and after (on the west side) the electron detector. The amount of vertical drift, $y(x)$, experienced by the positron/electron in the $\bar{E} \times \bar{B}$ field is given by the expression (Eq. 1)

$$y(x) = \frac{\Delta V}{d} \frac{L}{B} \sqrt{\frac{m}{2(E_K - qV(x))}} \qquad (1)$$

The electric field is assumed to be along the $x$ direction and the axial transport magnetic field is assumed to be along the $z$ direction in the above equation. Here, $\Delta V$ is the potential difference between the rectangular plates, $d$ is the distance between the plates, and $L$ the length of the plates. $B$ is the magnitude of the axial magnetic field at the rectangular plates, m is the mass of electron/positron, $E_K$ is the kinetic energy of the positron/electron when it enters the $\bar{E} \times \bar{B}$ field, and $q$ is the charge of the electron/positron. $V(x)$ is the potential experienced by the electron/positron at position $x$ in between the plates; assuming that $x = 0$ is at the plate that is biased to a positive potential $V_1$, $V(x) = \frac{V_2 - V_1}{d} x + V_1$. The values of the potentials applied on the plates are given in Table 2 for the two different positron energies obtained using the two different RGM settings (Table 1). The RGM is also a source of high intensity, low energy electrons [28] that are emitted following the implantation of fast positrons on the source and moderator, and by the irradiation of high atomic number UHV components by annihilation gamma rays. Some of these low energy electrons survive the magnetic 'energy filtering' and enter the positron beam system. The presence of these low energy electrons increases the accidental background to the electron spectrum from the sample as measured by the ToF spectrometer. In order to prevent the low energy electrons from entering the ToF spectrometer, the $\bar{E} \times \bar{B}$ plates on the east (on the source side), are biased such that there is a net negative potential at the center of the plates. The magnetic fields at the $\bar{E} \times \bar{B}$ plates have also been individually optimized to maximize the positron transport around the box in which the electron detector is housed.

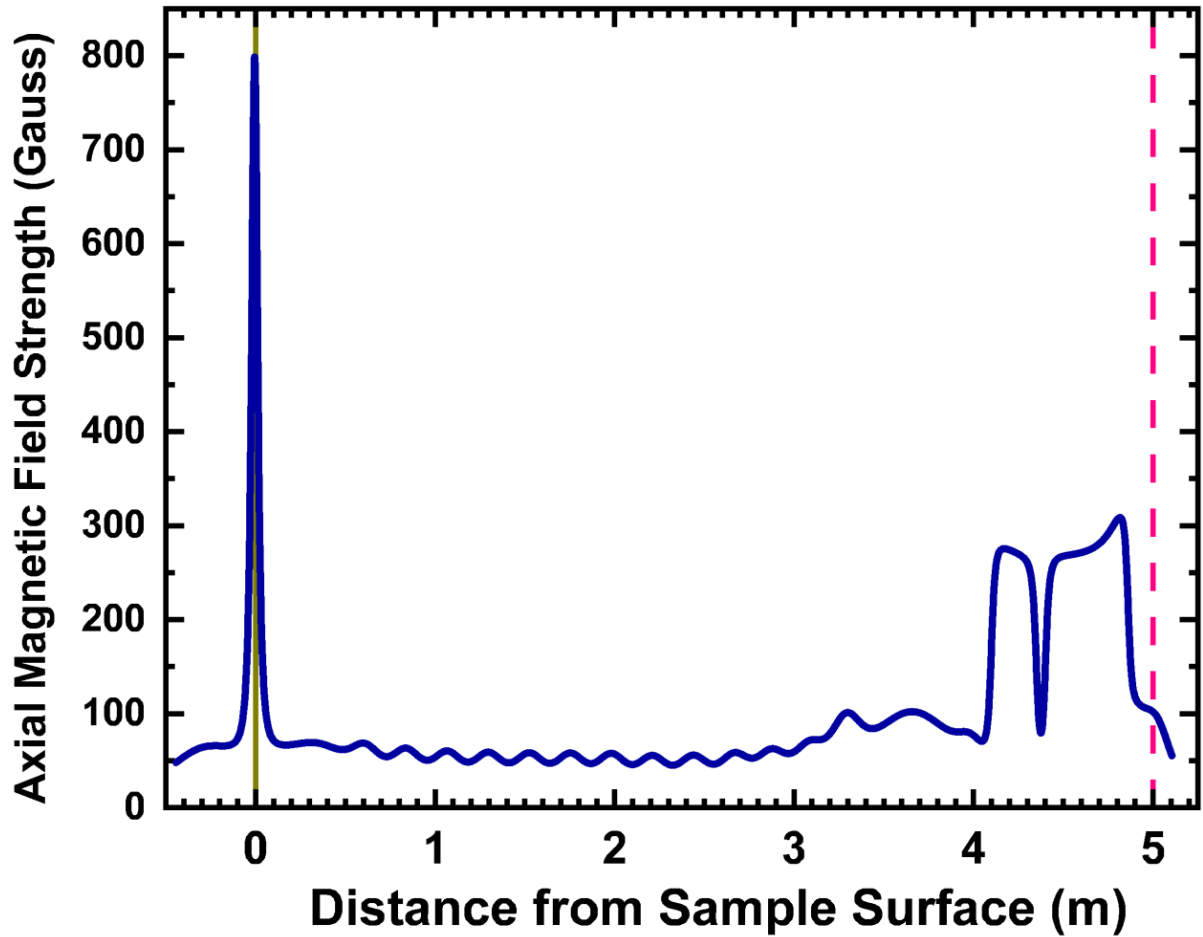

**Figure 3**. Axial magnetic field strength along the beam axis as a function of the distance from the sample. The axial field is achieved using a set of Helmholtz coils. The coil positions are shown in Figure 1 (a) and Figure 2. The dotted pink line shows the source position whereas the solid gold line shows the sample position. A permanent magnet near the sample creates a magnetic field gradient that adiabatically compresses the positron beam to ~ 2 mm. The magnetic field gradient also helps to parallelize the electron trajectories.

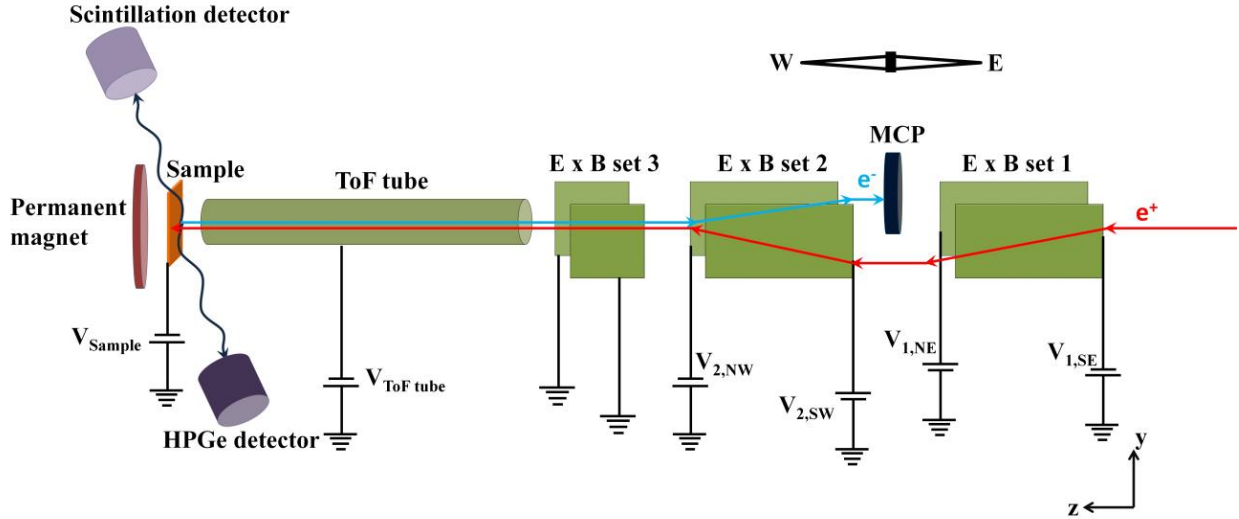

**Figure 4**. Schematic of beam optics of the advanced positron beam system. Positrons from the source are bent around the electron detector (MCP) using the $\bar{E} \times \bar{B}$ system. The $\bar{E} \times \bar{B}$ system consists of three sets of rectangular plates that can be biased to produce electric field along the x-direction, which when combined with the axial magnetic field along the z direction produces the required $\bar{E} \times \bar{B}$ field. In the present experiment one pair of the $\bar{E} \times \bar{B}$ plates (set3) is grounded. The $\bar{E} \times \bar{B}$ set 1 (east of MCP) causes the positrons to drift down in the y-direction whereas $\bar{E} \times \bar{B}$ set 2 (west of MCP) drifts the positron and the electrons up. The values of the potentials applied on the $\bar{E} \times \bar{B}$ plates are given in Table 2. The potentials and the axial magnetic field are optimized to maximize the positron counts at the sample.

### 2.3. Field-free ToF tube and positron beam energy

The slow positrons that are transported around the electron detector traverse a 3 m field-free ToF tube before entering the sample chamber. The ToF tube can be used as a retarding field analyzer to measure the energy of the positrons emitted from the RGM. For the retarding field energy analysis, the annihilation gamma photons at the sample were measured as a function of the magnitude of the positive potential applied on the ToF tube (Figure 5) using a high purity Ge detector placed nearby. The negative derivative of the retarding field analysis data gives the distribution of $p_z^2$ of positrons that reach the tube, where $p_z$ is the momentum of positrons parallel to the beam axis. The data shown in Figure 5 (a) is obtained when the RGM is operated at the low energy settings and the data in Figure 5 (b) is obtained when RGM is operated at the high energy settings. The kinetic energy of the positrons ($p_z^2/2m$) was measured immediately after the growth of the moderator film. With a moderator bias of 3 V (low energy settings), the peak of the positron kinetic energy distribution is at 4 eV, whereas the peak is at 15.7 eV when the moderator was biased to 15 V (high energy settings). The main drawback of the solidified rare gas moderator system is the relatively large spread in the energies of the slow positrons produced by the moderation process. In the low energy settings, the full width at half maximum (FWHM) of the energy distribution of positrons is 2 eV, consistent with previous reports [25, 26]; in the case of high energy settings the FWHM of the distribution is 1.25 eV.

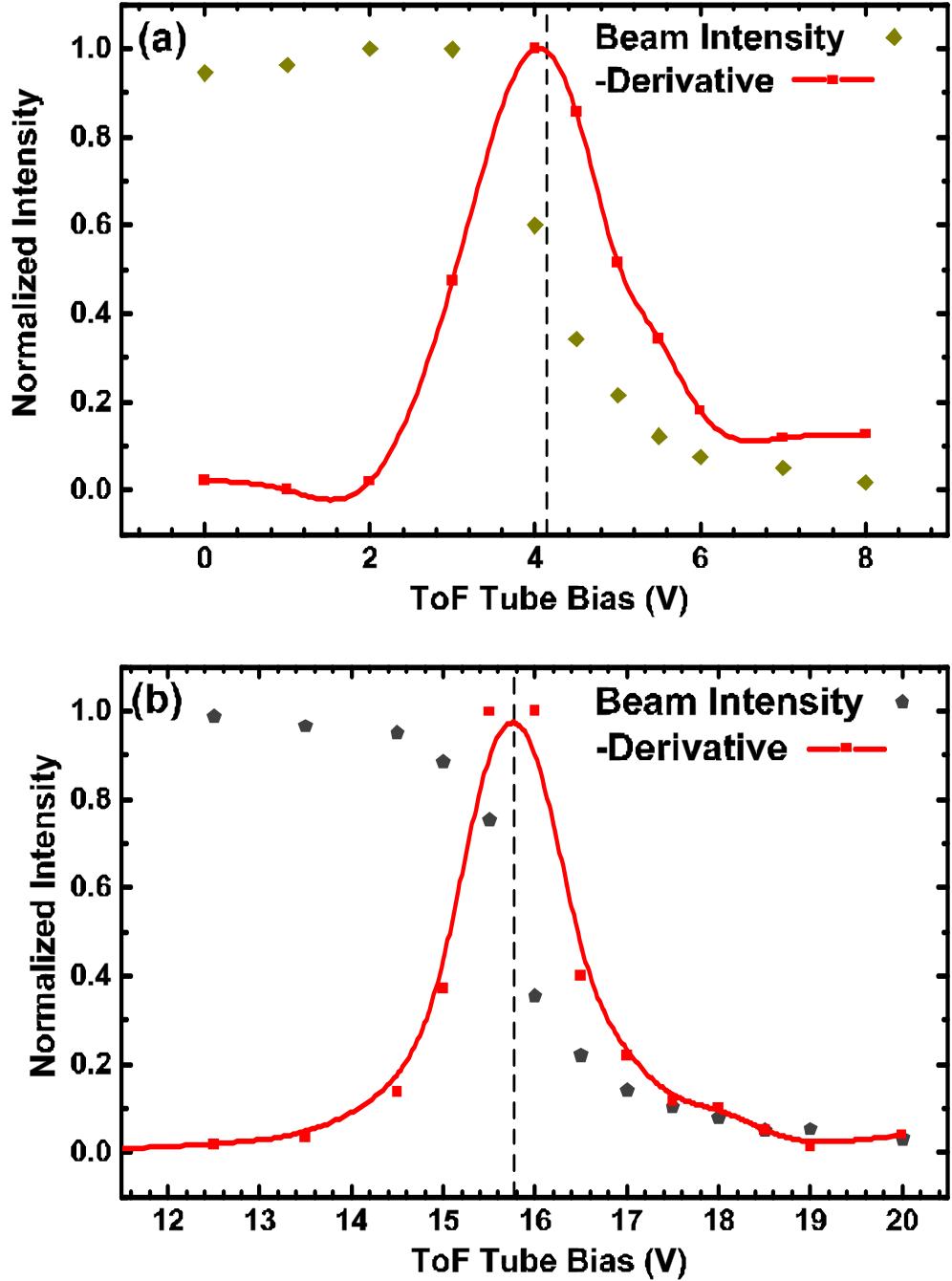

**Figure 5**. Kinetic energy ($p_z^2/2m$) distribution of positrons measured at the ToF tube for two different RGM settings. (a) Positron counts measured at the sample (solid diamond) as a function of the positive potential applied on the ToF tube with 'low energy setting' on the RGM. The negative derivative of the positron counts taken with respect to the ToF bias is represented by solid rectangles. The solid line through the derivative spectrum is an aid to the eye. The positron energy distribution has a peak at ~ 4 eV and has an FWHM of ~ 2 eV (b) Positron count rate and the corresponding negative derivative measured at the sample with 'high energy settings' on the RGM. The positron energy distribution has a peak at ~ 15.7 eV with a FWHM of ~ 1.25 eV.

## 2.4. Positron acceleration and the sample chamber.

The positrons that enter the sample chamber are further accelerated or decelerated by the bias applied on the sample with respect to the ground. This bias is applied using a high voltage (HV) vacuum feed-through, allowing the sample to be biased to a maximum of 20 kV. The positron beam energy can thus be continuously varied from ~ 0 eV to 20 keV by appropriately biasing the sample. The resulting mean kinetic energy of the positrons ($E_{K,p}$) on the sample is given as (Eq. 2):

$$E_{K,p} = q(V_{moderator} - V_{sample}) + E_{moderator} \qquad (2)$$

Here, $q$ is the charge of the positron, $V_{moderator}$ is the potential applied to the moderator, and $V_{sample}$ is the potential applied to the sample with respect to the ground. $E_{moderator} \approx 1$ eV [25, 26] is the mean energy of moderated slow positrons emitted from Ne moderator.

The sample, along with the permanent magnet, is mounted on a manipulator with three degrees of linear freedom (East/West, North/South, Up/Down) and two degrees of rotational freedom (roll and yaw). The corresponding mounting arrangement is illustrated in Figure 1(b) and allows for two samples to be installed simultaneously. The sample to be analyzed by the positron beam is brought to the analysis position by rotation about the z-axis (roll). The rotation and linear motion allow the sample to be transported to the sputtering position of a sputter ion gun (Perkin Elmer Φ 04-191); to the analysis positions of a cylindrical Auger electron optics apparatus (Perkin Elmer Φ 10-155); or to the analysis positions of low energy electron diffraction (LEED) optics (Perkin Elmer Φ 15-120), all of which are attached to the sample chamber. The linear motion is also used to make contact with the vacuum side pin of the HV feed-through. A leak valve is also attached to the sample chamber for controlled gas exposure. The sample chamber is equipped with two anti-parallel detector wells; one of which houses the high purity germanium (HPGe) detector and the other a fast timing scintillation detector ($BaF_2$ or NaI, depending on the experiment being performed).

## 2.5. Annihilation gamma energy spectroscopy

The energy spectrum of the annihilation gamma photons is derived from the signals produced coincidentally by the HPGe detector (GEM-10195, 7.2% relative efficiency at 1.33 MeV) and the scintillation detector placed inside the detector wells. A schematic of the semi-analog electronics system used for the collection of signals from the HPGe detector is shown in Figure 6. If the requirement for a gate input is activated in the multichannel analyzer (MCA), then only those HPGe signals which are in coincidence with the scintillation detector signals are stored. Set up in this way, the energy resolution of the semi-analog HPGe gamma spectrometer at 511 keV is $1.08 \pm 0.05$ keV.

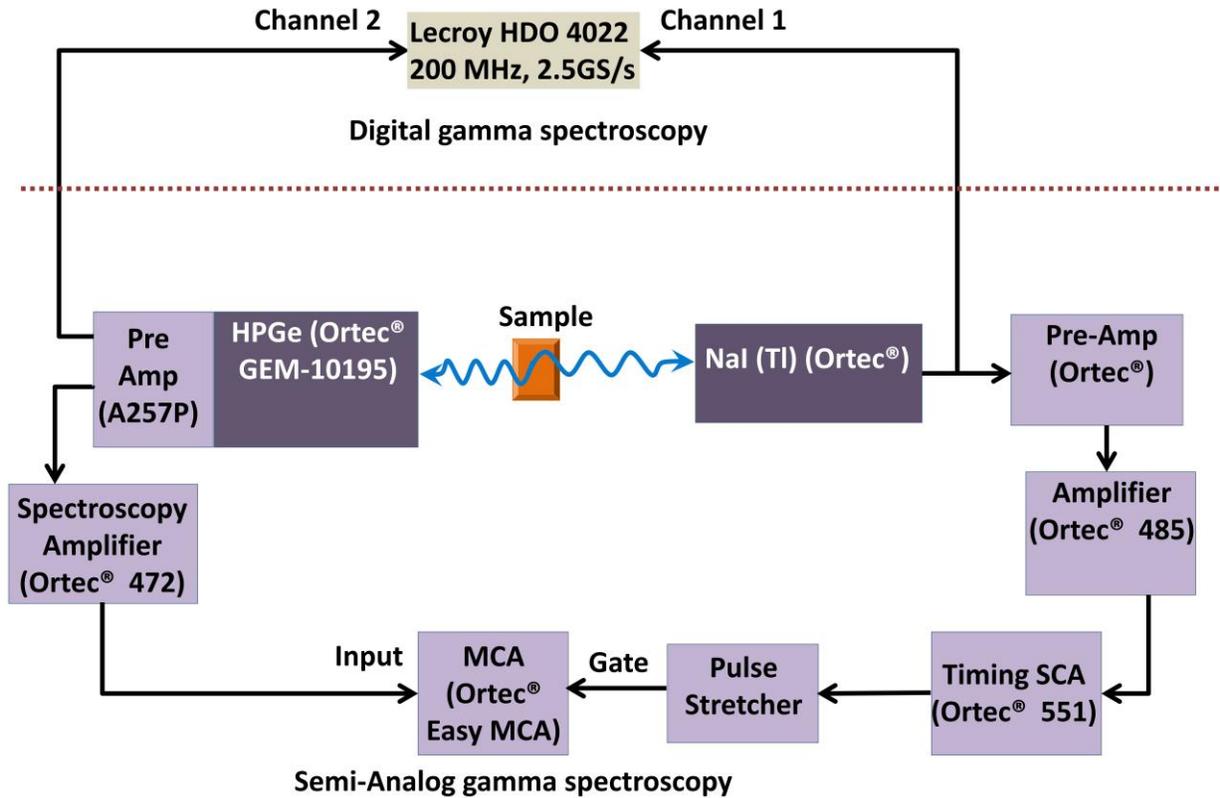

**Figure 6**. Schematic of the digital and semi-analog data acquisition system for gamma energy spectroscopy. In the semi-analog setup, the signals from the HPGe detector are Gaussian shaped and amplified using a spectroscopy amplifier. The shaping time is chosen to be 6 μs. If the requirement for a Gate signal is activated in the MCA, then HPGe signals are collected in coincidence with the gamma detected by the NaI detector. In digital gamma spectroscopy, the raw signals from the detectors are fed to a 12-bit, 2.5 GS/s Lecroy high definition oscilloscope (HDO). The measurement can be carried out with or without the requirement of two-gamma coincidence. The pulses are post-processed with the algorithm described in [29] to build the gamma energy histogram.

The advanced positron beam system is also equipped with a digital data acquisition system, where the outputs from the detectors are recorded directly. A second pre-amplifier output from the HPGe detector is fed to one of the channels of a 12-bit, two-channel LeCroy® HDO 4022 oscilloscope; if two gamma coincidences are required, then the raw, unamplified signal from the NaI scintillation detector anode is fed to the second channel of the oscilloscope. Analysis software developed by Gladen et al. [29] extracts the energy information from the detector signals with an optional, user-selected coincidence condition. The energy resolution of this digital HPGe gamma spectrometer at 511 keV is 1.39 ± 0.04 keV using the digital implementation of the conventional pulse shaping algorithms.

### 2.6. *The ToF electron spectrometer*

Electrons that are emitted from the sample as a result of positron implantation or annihilation spend most of their flight time in a field-free ToF tube prior to their entrance into the $\bar{E} \times \bar{B}$ system. The field-free ToF tube provides to the electron a path free of any stray electric fields, conserving the component of the electron velocity parallel to the beam axis. It is also possible to apply a negative potential to the ToF tube, with respect to the sample, to reflect electrons with insufficient energy to overcome this applied bias. Additionally, the negative bias decelerates the electrons to a lower energy that persists during their traversal time through the ToF tube, resulting in improved resolution in the determination of the energy of the detected electrons [30-32]. After exiting the ToF tube, the electrons enter the $\bar{E} \times \bar{B}$ system, where they are drifted upwards to the entrance of the electron detector according to equation 1. The active area of the MCP presently in the advanced positron beam system is ~ 42 mm. The higher energy electrons will hardly undergo a vertical drift with applied fields in the $\bar{E} \times \bar{B}$ system, whereas the lower energy electrons will experience a much more significant drift. Hence, well-determined $\bar{E} \times \bar{B}$ settings are required to maximize the range of electron energies that make it to the active area of the MCP electron detector. Additionally, the length of the $\bar{E} \times \bar{B}$ system (0.36 m) is much smaller than the total flight path of the electrons (~ 3 m); therefore, this region does not contribute significantly to the total ToF.

### 2.6.1. *Principle of operation of ToF spectrometer*

The ToF of the electron is obtained from the time difference between the signal produced by the MCP upon electron detection and the signal produced the gamma detector upon the detection of the annihilation gamma. In short, the ToF spectrometer determines only the kinetic energy associated with $v_z$, the velocity component associated with the motion along the z-axis (beam axis). An accurate determination of the kinetic energy of electrons emitted from the sample depends on having ideally no component or, at worst, a very small component normal to the axial velocity. This is achieved by using a magnetic field parallelizer [10]. The permanent magnet placed just behind the sample produces an axial magnetic field gradient (as seen in Figure 3) that reduces the angular divergence of the electrons that enter the ToF analyzer. If $v_{ns}$ is the component of the electron velocity normal to the z-axis at the sample, $v_{zs}$ is the velocity component parallel to the z-axis at the sample, $v_{na}$ is the component of the electron velocity normal to the z-axis at the analyzer (the region with reduced magnetic field), and $v_{za}$ is the velocity component parallel to the z-axis at the analyzer, then by energy conservation:

$$v_{total}^2 = v_{ns}^2 + v_{zs}^2 = v_{na}^2 + v_{za}^2. \quad (3.a)$$

Here, $v_{total}$ is the total velocity of the electron with mass $m$. By adiabatic invariance [27]:

$$\frac{v_{ns}^2}{B_s} = \frac{v_{na}^2}{B_a}. \quad (3.b)$$

Here, $B_s$ is the axial magnetic field at the sample and $B_a$ is the magnetic field at analyzer. By an application of Equations 3.a and 3.b, we arrive at Eq. 4:

$$\sin\theta' = \left(\frac{B_a}{B_s}\right)^{1/2} \sin\theta \quad (4)$$

Here, $\theta$ is the polar angle at which the electron is emitted from the sample with respect to the z-axis and $\theta'$ is the polar angle at which the electron enters the analyzer. For $\theta = \pi/2$, the maximum angle at which the electron can be emitted from the sample, the corresponding maximum angle at which the electron enters the analyzer is: $\theta'_{max} = \arcsin\left(\left(\frac{B_a}{B_s}\right)^{1/2}\right)$. In this case, an electron with kinetic energy $E_k = \frac{1}{2}mv_{total}^2$ at the sample and emitted at $\theta = \pi/2$, will be measured as an electron with energy $E_k = \frac{1}{2}mv_{total}^2 \cos^2\theta'_{max}$ by the ToF analyzer. The fractional uncertainty in the determination of the energy due to the angular divergence of the electron emission is thus:

$$\left(\frac{\delta E_k}{E_k}\right)_{angular\ divergence} = \sin^2\theta'_{max} \cong \frac{B_a}{B_s} \quad (5).$$

For the present apparatus, $\frac{B_a}{B_s} \cong \frac{1}{15}$, and thus $v_{na}^2 \cong v_{ns}^2/15$, greatly reducing the normal velocity component of the electron entering the ToF analyzer. The magnetic parallelization occurs in its entirety within 10 cm of the sample surface; a negligible distance when compared to the total flight path of the electron. Hence, the variance in the amount of time taken for parallelization to occur for electrons with various energies and various angular divergences is negligible and is not considered for the total timing uncertainty.

The preceding discussion implies that the amount of time the electron spends in the magnetic field parallelizer and the $\bar{E} \times \bar{B}$ system is markedly smaller than the time the electron spends in the field-free ToF tube; therefore, since the electron spends most of its flight time in a field-free environment, we can approximately express the electron kinetic energy, as measured by the ToF spectrometer, as $E_{K,e} = \frac{1}{2}m\frac{L^2}{t^2}$, where $t$ is the measured ToF of the electron and $L$ is the length of the electron flight path. The fractional error in energy due to the uncertainty in the determination of the ToF ($\delta t$) from the electron and gamma signals is accordingly given by:

$$\left(\frac{\delta E_k}{E_k}\right)_{time\ uncertainity} \cong \frac{2\delta t}{t} = \frac{2\delta t}{\sqrt{0.5mL^2}}\sqrt{E_k}. \quad (6)$$

$\delta t$ contains within it the timing resolution of the MCP electron detector, the timing resolution of the gamma detector, and the resolution of the associated timing electronics or algorithms. If we assume that the error in the determination of the electron energy due to the angular divergence and the error due to timing uncertainty are uncorrelated, then we can express the total error, or the instrumental broadening, introduced by the ToF spectrometer as the summation in quadrature of the individual errors, yielding:

$$\left(\frac{\delta E_k}{E_k}\right)_{spectrometer} = \sqrt{\frac{(2\delta t)^2}{0.5mL^2}E_k + \left(\frac{B_a}{B_s}\right)^2} \quad (7)$$

The energy resolution of the ToF spectrometer can, therefore, be improved by enhancing the timing resolution of the detectors involved and by increasing the magnetic field gradient between the sample and the analysis position. There is however a limit to the amount by which one can increase $B_s$ or decrease $B_a$, as either approach will negatively affect the positron count rate due to positron reflection at the sample. We can improve the energy resolution by increasing the flight path length, $L$, in order to increase the separation, in time, of the electrons to a value great than the time resolution of each detector. The energy resolution can be further improved by applying a negative potential on the field-free tube with respect to the sample. The negative potential decelerates the electron, increases the ToF, $t$, and thus improves the energy resolution, as can be seen in Equation 6 and as demonstrated by Ohdaira et al. and Hugenschmidt et al. [30-32].

Finally, if $\Delta E_{k,intrinsic}$ is the intrinsic width of a peak (as measured by an ideal spectrometer) in the electron spectrum, then the total width of the peak as measured by a ToF spectrometer will be

$$\left(\frac{\delta E_k}{E_k}\right)_{total} \cong \sqrt{\frac{(2\delta t)^2}{0.5mL^2}E_k + \left(\frac{B_a}{B_s}\right)^2 + \frac{(\delta E_k)^2_{intrinsic}}{E_k^2}} \quad (8)$$

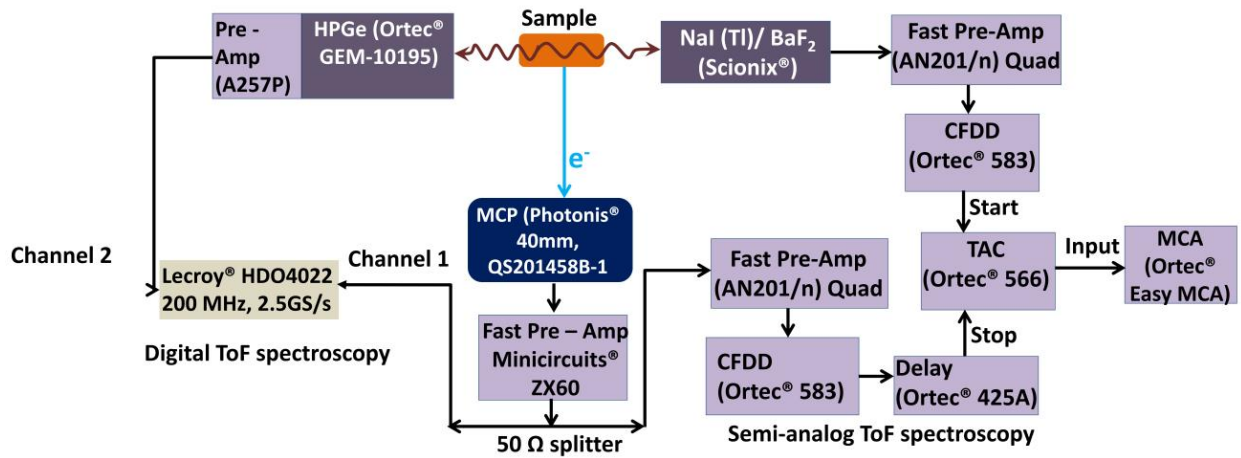

**Figure 7**. Schematic of the digital and semi-analog data acquisition system for ToF spectroscopy. In the semi-analog setup, the signal from the scintillation detector (after amplification) is fed to a CFDD to produce the time-pick off signal for the TAC. The time pick-off signal produced from the MCP output by a second CFDD stops the TAC, and the resulting TAC output is used by the MCA to build the ToF histogram. The pre-amplified output signals from the HPGe and MCP are fed directly to the Lecroy oscilloscope in digital ToF spectroscopy. The ToF histogram is constructed using the algorithm detailed in [29]. The output from the fast scintillation detector can be used instead of the HPGe signal to construct the digital ToF spectra.

### 2.6.2. *Data collection*

The apparatus is equipped with a semi-analog timing system and a digital data acquisition system to construct an analog or digital ToF spectra. A schematic of the semi-analog timing system is shown in Figure 7. The fast pre-amplified signal from the scintillator detector is fed to the constant fraction differential discriminator (CFDD), which produces a time pick-off signal. The signal from CFDD starts the charging of the capacitor in a time-to-amplitude converter (TAC). Likewise, the fast pre-amplified MCP signal is fed to another CFDD that produces a corresponding time pick-off signal. If the MCP signal occurs within the 5 μs coincidence window set on TAC, then the charging of the capacitor is stopped. The potential developed across the capacitor is then directly proportional to the time difference between the two signals. The voltage of the subsequent pulse produced by the TAC is used by the MCA to construct a time histogram.

In the digital ToF PAES system, the pre-amplified signal from MCP is fed to one of the channels of the oscilloscope and the signal from the gamma detector (HPGe or NaI or $BaF_2$) is connected to the second channel. One of the major benefits of the digital ToF system is the ability to use the HPGe detectors to derive the time of arrival of the gamma signal. HPGe detectors have superior energy resolution but are known to suffer from the relatively poor timing resolution that results from the variable rise times of the leading edge of the pre-amplified signals. These rise times depend primarily on the physical position of the electron-hole pair creation in the detector [33]. The software, as described in Ref. [29], takes these factors into account and, using an "extrapolation of leading-edge timing" (ELET) algorithm, derives the time of arrival of the gamma pulses. The software was developed such that it is capable of accurately determining the time pick-off signal from either the HPGe detector or the fast timing scintillation detectors, depending on the experiment.

The use of an HPGe detector for timing allows the correlation of the time-of-flight of the electron to the energy of the annihilation gamma (determined from the amplitude of the HPGe signal) without the need of any further complex coincidence conditions. This step allows us to develop a two-dimensional matrix of correlated electron ToFs and gamma energies. A careful analysis of the two-dimensional data yields the spectrum of the Doppler-broadened annihilation gamma photo-peak (measured in high resolution using HPGe) corresponding to specific Auger electron peaks in the ToF spectrum. If a fast timing detector was used for timing in conjunction with HPGe for energy determination, a triple coincidence condition would need to be invoked between the MCP signal and the two gamma detector signals to generate the two-dimensional electron ToF - gamma energy correlation matrix. This condition would significantly reduce the count rate of the experiment and increase the time required for measurements. The requirement that two annihilation gammas be detected at two detectors which are 180 degrees apart will also preclude the measurement of three-gamma emission due to ortho-positronium (o-Ps) annihilations. Ps formation may be associated with electron emission through processes such as the Auger-mediated neutralization of the positron. The final advantage of using HPGe for timing that will be remarked upon here is that the ToF and gamma energy spectrum can be derived using a two-channel digitizer/oscilloscope.

### 2.6.3. *Time resolution of the ToF spectrometer*

The time resolution ($\delta t$) of the present ToF spectrometer can be estimated by applying equation (8) to secondary electron spectra collected with higher sample biases (> 100 V). When an appropriately large sample bias is applied, the uncertainty in the determination of the energy of the electron due to the angular divergence of electron emission will be negligible compared to both the timing uncertainty and the inherent width of the secondary electron spectrum. The reasons for the reduced contribution of the magnetic field gradient component to spectrum broadening are: (i) the applied negative bias on the sample with respect to the ToF tube creates an electric field parallel to the beam axis which results in electron emission with much larger $v_{zs}$ (the velocity component parallel to the z-axis at the sample); (ii) positron-induced secondary electrons can be emitted from a few atomic layers below the surface and therefore the angular distribution of electron emission is less isotropic and more directed towards the beam axis [34]. One more fact that aids in the determination of the timing resolution is that the inherent FWHM of the secondary electron spectra is approximately equal for large (> 100 eV) incident positron/electron energies [35]. With these approximations, Equation (8) will have only two unknowns and, by setting up two equations, both the timing resolution and the inherent width of the secondary electron spectra can be determined. To achieve this, the FWHM of the energy spectra of secondary electrons collected with sample biases of -100 V and -150 V was used. The inherent width of the secondary spectra for an incident positron energy of 104 eV or 154 eV was determined to be $9.5 \pm 1.5$ eV, a value consistent with the width of the secondary spectra reported in literature [35]. The time resolution of the ToF spectrometer when the MCP is used in conjunction with the NaI(Tl) scintillation detector is $23 \pm 3.7$ ns, and the time resolution when using the MCP-HPGe detector combination is $22 \pm 5$ ns. The digital analysis of HPGe signals provides a timing resolution comparable to that acquired when using a fast scintillation detector like NaI(Tl); a result which allowed us to use the HPGe detector for the coincident measurement of the energy of the positron-induced electrons and the energy of the associated annihilation-induced gamma ray.

### 2.6.4. *Energy resolution of the ToF spectrometer*

As shown in Equation (8), the energy resolution of the ToF spectrometer is dependent on the measured energy and other parameters: the timing uncertainty, the magnetic field gradient, and the electric fields in the path of the electron trajectory. The contribution of the angular divergence to the total energy resolution is also critically dependent on the applied bias on the sample, as noted in the preceding section, which makes analytical prediction of the instrument response difficult. In order to accurately predict the instrument response and, thus, the energy resolution of the ToF spectrometer with the electromagnetic field settings applied during the experiment, the advanced positron beam system was simulated using SIMION 8.1 [36]. The mechanical assembly of the electromagnetic optics was accurately ($\pm$ 1 mm) built in SIMION and the trajectory of electrons emitted from the sample were simulated for the low energy electromagnetic settings given in Table 1 and Table 2.

In order to determine the energy resolution of the ToF spectrometer at experimental conditions, mono-energetic electrons were flown from the sample, biased to -1 V. The electrons were generated randomly on a 2.0 mm diameter circle and emitted with isotropic angular distribution. The physical region of electron emission from the sample was determined by flying positrons from the source to the sample and recording the incident positron distribution on the sample surface. The ToFs of the mono-energetic electrons reaching the MCP were recorded and the corresponding ToF histogram was generated with a bin width equal to that of the experimental spectrum. The ToF spectra thus generated for each electron energy were further convoluted with a Gaussian that has FWHM of 22 ns to take into account the broadening due to the timing uncertainty of the detectors and the corresponding electronics.

The simulated ToF spectra were converted to energy using a simulated calibration curve utilizing a procedure equivalent to the one employed for the experimental calibration curve. An electron with zero kinetic energy was flown from the center of the sample at different sample biases and with a constant ToF tube bias of -0.5 V. The inverse of the square root of the sample bias is then expressed as a function of the respective ToF of the 0 eV electron. The simulated calibration plot is fit to a fourth-order polynomial and is coupled with an appropriate Jacobian to convert the ToF spectrum into its corresponding energy spectrum. The FWHM of the energy spectrum of the mono-energetic electrons thus produced was used to generate the energy resolution plot shown in Figure 8. The energy resolution is better than 20% of the measured energy ranging from 5 eV to 500 eV, which is appropriate for positron annihilation induced Auger electron spectrum. The best resolution of the spectrometer is at 50 eV and is approximately twice that which was obtained using a 1 m ToF spectrometer with ~ 2 ns timing resolution [11]. It should be noted that the 1 m ToF spectrometer was equipped with a fast $BaF_2$ detector that was responsible for the ~ 2 ns timing resolution [11]. In the present case, even though a detector with relatively poor timing resolution was employed (~ 10 times worse), the corresponding influence on the total energy resolution was reduced by the three-fold increase in the total flight path and, further, by the use of a larger magnetic field gradient at the sample.

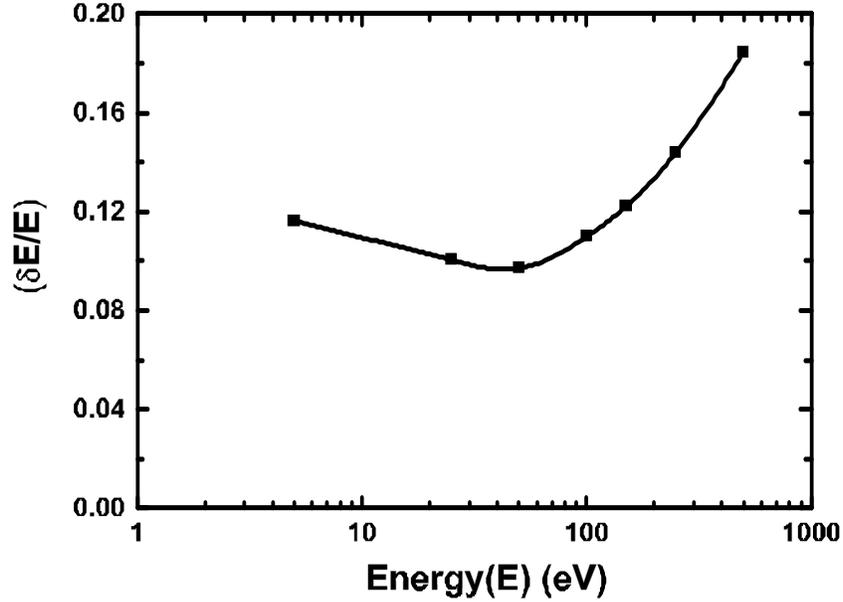

**Figure 8**. Energy resolution ($\frac{\delta E}{E}$) of the 3 m ToF spectrometer as a function of the measured kinetic energy ($E$) of the positron-induced electrons. The relative energy width was determined by simulating the trajectory of monoenergetic electrons through the 3 m ToF spectrometer in SIMION [36]. The settings of the electromagnetic optics used for the simulation is equal to those shown for the "low-energy-settings" in Table 1 and Table 2. The solid line through the points is an aid to the eye. The relative energy width is smaller than 20% over the entire energy range of interest even with the employment of an HPGe detector for measuring the time of arrival of the annihilation gamma photon. The energy resolution is appropriate for the determination of the ToF of positron-induced electrons. The increased flight path and the larger magnetic field gradient allowed us to keep the energy resolution below 20% of the measured energy. The ability to determine the energy of electrons with less than 20% error along with the energy of the Doppler-shifted annihilation gamma photon with high energy resolution was important for measurements described here.

### 2.6.5. *Energy calibration of the ToF spectrometer*

The experimentally measured time difference between the annihilation gamma photon, detected by the HPGe detector, and the positron-induced electron, detected by the MCP, is converted to energy using an experimental calibration plot. Since there are regions in the flight path that are not field-free, the functional relation between the ToF and energy is more complex than $E_{K,e} = \frac{1}{2}m\frac{L^2}{t^2}$; the exact relation is found experimentally by collecting secondary electron spectra at various sample biases. A few representative ToF spectra of secondary electrons obtained with different biases on the sample are shown in Figure 9(a). The spectra were collected with the low energy settings given in Table 1 and Table 2. As the sample bias is increased, the secondary spectra shift to lower times due to the increased kinetic energy of the electrons emitted from the sample. The minimum kinetic energy of electrons in a secondary electron energy spectrum is equal to $-eV_{sample}$, where $V_{sample}$ is the bias on the sample. The experimentally measured ToF of the electron that travels with energy equal to $-eV_{sample}$ through a majority of the 3 m flight path is identified from the secondary spectrum as the starting point of the rising edge of the spectrum, illustrated by the vertical line in Figure 9(a). The inverse of the square root of the sample bias is then expressed as a function of the ToF of the minimum energy electrons, as shown in Figure 9(b). It should be noted that the experimentally measured ToF and the simulated (true) ToF are separated by only a constant time delay that is representative of the unknown delays associated with the detection and acquisition of the electron and gamma signals. Their equivalence is shown in Figure 9(b), where the simulated energy calibration points are provided along with the experimentally determined calibration curve which has been shifted by 15 ns. The experimental calibration plot is fit to a fourth order polynomial and with an appropriate Jacobian the ToF spectra is converted to energy spectra.

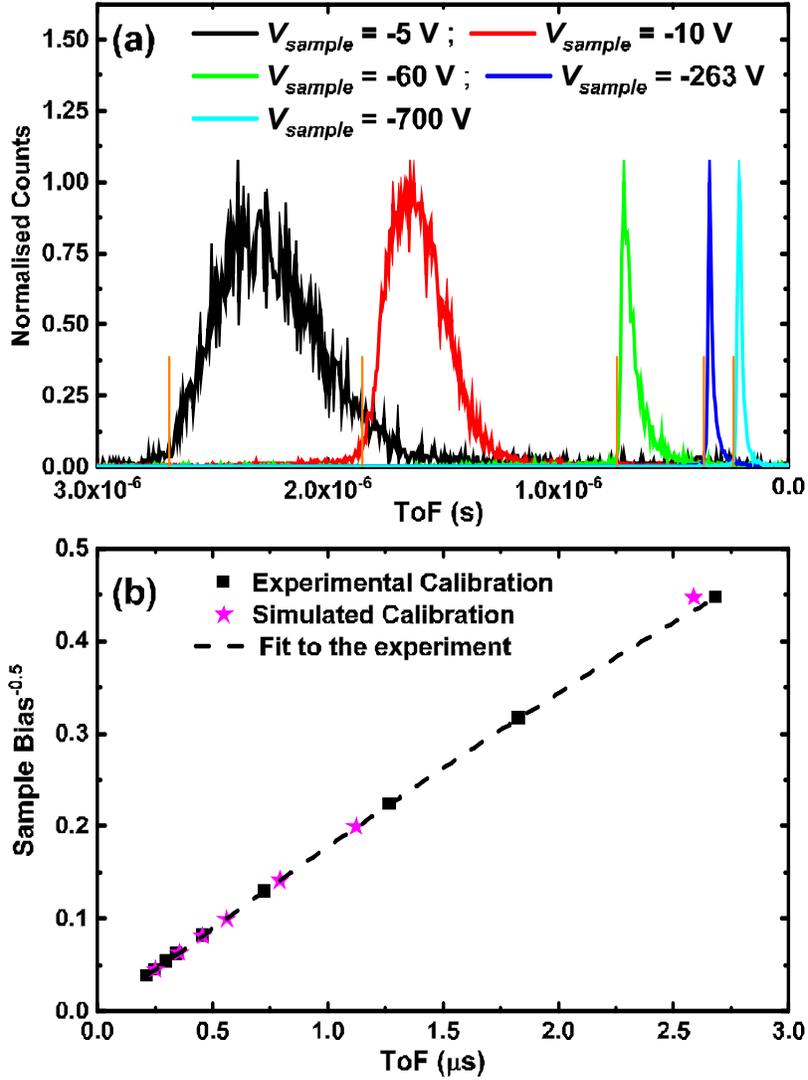

**Figure 9**. (a) Peak-normalized ToF spectra of secondary electrons collected with different biases on the sample (bilayer graphene on Cu). The energy of the outgoing electrons increases with increasing negative bias on the sample. This is seen by the movement of the secondary peak to lower ToFs. The edge of the secondary peak, shown here by a short vertical orange line, represents electrons which escaped the sample surface with 0 eV. These 0 eV electrons are accelerated by the sample bias and travel with a kinetic energy equal to $-e\,V_{sample}$ through the ToF tube. Since the total flight time is dominated by the flight time through the field-free ToF tube, the measured ToF of the electron can be considered to be $\propto V_{sample}^{-0.5}$. (b) The relationship between the measured ToF and the kinetic energy of the electron has been derived by finding the ToF corresponding to the zero eV electrons in the secondary electron spectra and expressing $1/\sqrt{V_{sample}}$ as a function of the zero eV electron ToFs. The dashed line is a fit to the calibration plot with a fourth order polynomial. The functional relationship thus derived was used along with an appropriate Jacobian to convert the ToF spectrum to an energy spectrum. The calibration points obtained through SIMION [36] simulation of the ToF spectrometer are also shown (magenta stars). The simulated and the experimental ToFs are separated by a constant delay (15 ns) which is due to the unknown delays of the experimental measurement of the ToF. We have corrected for this delay in our ToF spectra. The total calibration spectra were collected in two days.

## 3. RESULTS AND DISCUSSION

### 3.1. *Positron annihilation induced Auger electron spectrum*

The positron annihilation-induced Auger electron spectrum (PAES), measured using the advanced positron beam system from a sample of bilayer graphene on polycrystalline Cu substrate, is shown in Figures 10(a) and 10(b). The spectrum was measured with a sample bias of -1 V and at the low energy settings; the maximum energy of the positrons was ~ 6 eV (Figure 5(a)). The ToF spectrum constructed using the MCP - NaI(Tl) detector combination in the semi-analog mode, and the spectrum constructed using the MCP – HPGe detector combination using fully digital methods are shown together for comparison in Figure 10(a). A noticeable distinction between the two spectra is a considerable difference in statistics; this is due primarily to both the lower efficiency of the HPGe detector and the necessity of discarding ~ 20% of the collected HPGe detector pulses that are affected by various distortions [29]. The main features of the ToF spectrum are the Auger peaks corresponding to the Cu $M_{2,3}VV$ (58 eV), the C KVV (263 eV) and the O KVV (503 eV) Auger transitions. The spectrum after being converted to energy is shown in Figure 10(b). Apart from the main Auger peaks, there is significant intensity in the low energy region of the ToF spectrum due to secondary processes like Auger mediated positron sticking and positron impact induced secondary electron emission. The region below 30 eV also has appreciable contribution from low energy Auger emissions from carbon and oxygen [13, 37-38] and from the inelastically scattered high energy CVV Auger electrons. Note the most salient feature of the spectra comparison: the line shapes of the Auger peaks measured using the MCP – NaI(Tl) detector combination are fully captured by the digitally-processed ToF spectrum produced with the MCP - HPGe detector combination. The ability to use an HPGe detector for timing has been aided by: (i) the introduction of a relatively long flight path to increase the time-of-flight of all electrons within the detectable energy range to values significantly great than the typical timing resolution of the HPGe detector; and (ii) the implementation of digital signal processing [29] which allows for the selection of HPGe pulses with appropriate shapes for timing.

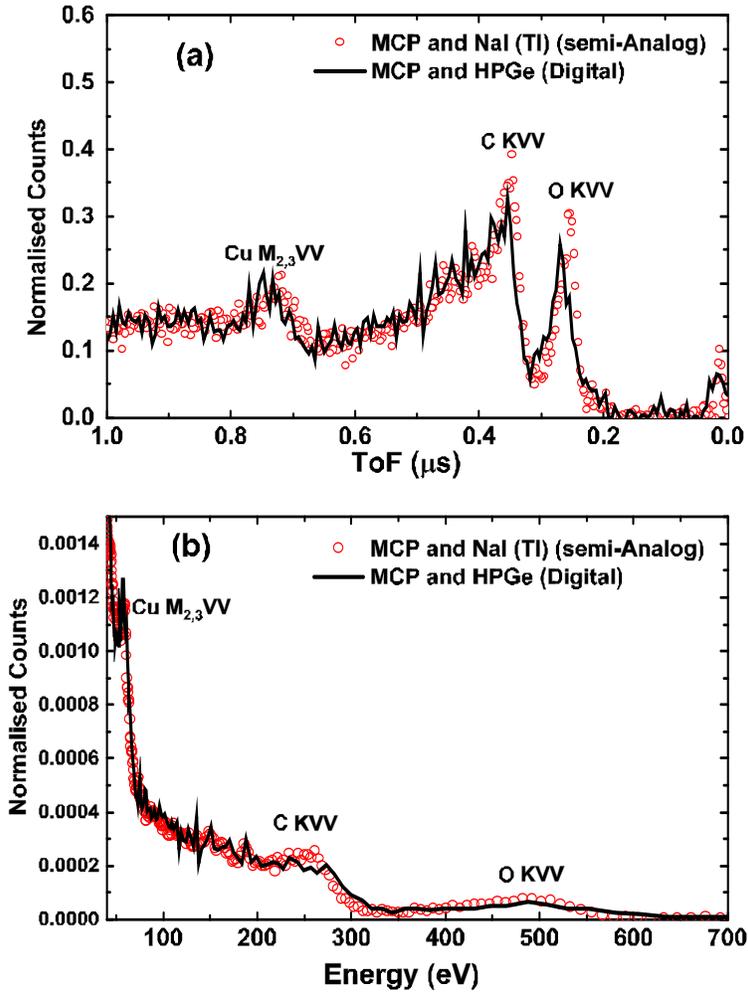

**Figure 10**. (a) ToF spectra of positron annihilation-induced Auger electrons from bilayer graphene on a Cu substrate. The spectrum collected using the semi-analog MCP-NaI(Tl) combination (red circles) is compared to the spectrum collected using the fully digitized MCP-HPGe combination. Both spectra have been area normalized with a constant background subtracted to correct for accidental coincidences ($5.8 \times 10^{-5}$ counts/(channel second) in case of MCP-NaI(Tl) combination and $1.6 \times 10^{-5}$ counts/(channel second) in case of MCP-HPGe combination). The spectrum using MCP-HPGe detectors fully captures the spectral features measured using an MCP and a fast scintillator detector. The major difference is the poor statistics of the spectra collected digitally, although both spectra were collected simultaneously for the same duration of time. This is because of the low efficiency of the HPGe detector (7.2% relative efficiency at 1.33 MeV) in comparison to NaI and the digital rejection of ~ 20% of the HPGe detector pulses (which did not satisfy the shape requirements). (b) The energy spectra of the positron annihilation-induced Auger electrons. The main features of the ToF spectrum are the Auger peaks corresponding to the Cu M2,3VV (58 eV), the C KVV (263 eV) and the O KVV (503 eV) Auger transitions. Apart from the main Auger peaks, there is significant intensity in the low energy region of the ToF spectrum due to secondary processes: Auger mediated positron sticking and positron impact-induced secondary electron emission. The energy region < 30 eV also has significant contribution from low energy Auger emissions from C and O and from the inelastic scattering of the higher energy CVV Auger electrons. With 6.9 MBq positron source and 0.2% moderation efficiency, a total collection time of ~ 30 days was required to produce these data.

*3.2    Doppler broadening spectrum*

The Doppler broadening spectra measured using the HPGe detector with different coincidence conditions are shown in Figure 11(a) and 11(b). The spectra of Figure 11(a) were produced by digital data acquisition and analysis techniques and were acquired with -900 V applied on the sample with respect to the ToF tube (the mean kinetic energy of the incident positrons was ~ 903 eV). The high-energy region of the Doppler broadened annihilation gamma spectrum, collected without any coincidence conditions (i.e. the "singles" spectrum), is compared with the Doppler broadening spectrum of gamma photons which were collected in coincidence with the positron-induced electrons as detected by the MCP (Figure 11(a)). The electron coincidence condition produces a marginal improvement in the high energy gamma background which can be attributed to the rejection of natural background sources. Figure 11(b) shows the high-energy region of the NaI(Tl) – HPGe coincidence Doppler broadened annihilation gamma spectrum (collected with -20 kV applied on the sample) compared to the singles spectrum. This singles spectrum is the output of the semi-analog gamma spectroscopy system; the coincidence spectrum has been constructed using the digital gamma spectroscopy system. The gamma-gamma coincidence condition results in an order of magnitude decrease in the high energy background due to the rejection of events where both 511 keV gamma (one of them after scattering) are detected at the same detector. This is comparable to the improvement achieved using other HPGe-NaI coincidence systems reported in literature [39].

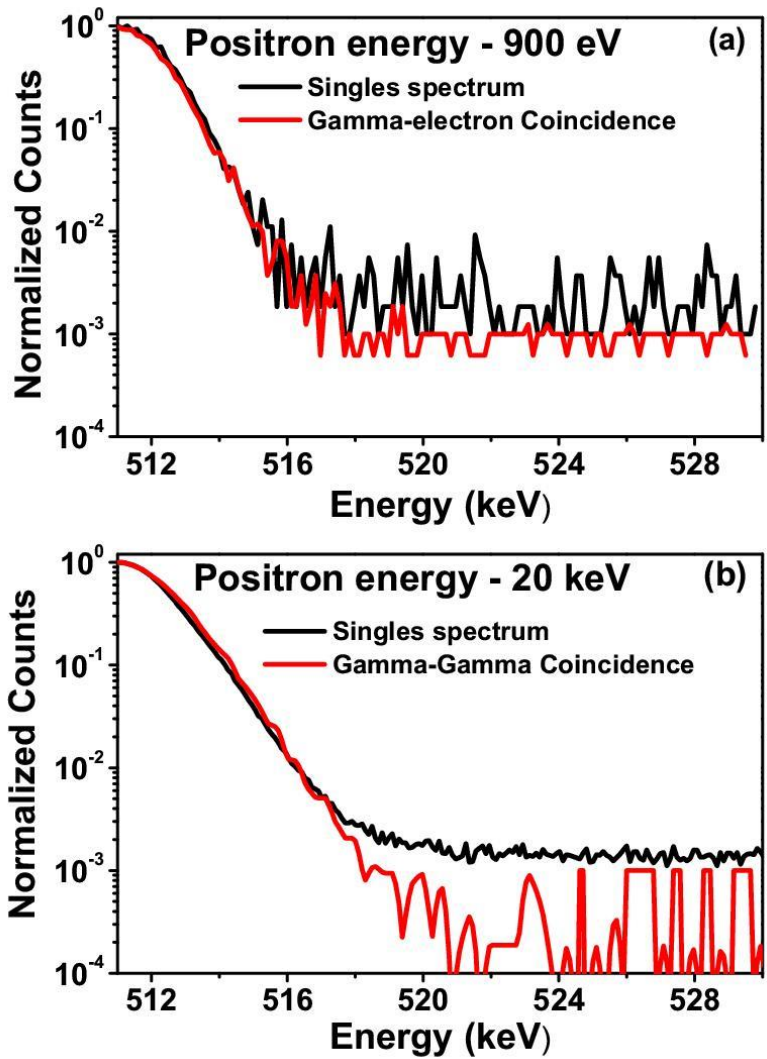

**Figure 11**. (a) Energy spectra of Doppler-shifted annihilation gamma photons measured using an HPGe detector with two different coincidence conditions. The sample (bilayer graphene on a polycrystalline Cu substrate) is biased to -900 V with respect to the ToF tube, resulting in a mean kinetic energy of the incident positrons is ~ 903 eV. The annihilation gamma spectrum measured without any coincidence condition ('singles spectrum') is compared to the annihilation gamma spectrum collected in coincidence with positron-induced electrons. The coincidence condition does not alter the overall line shape but improves the high energy background due to the rejection of ambient background counts. In this case both spectra were constructed using digital data acquisition and analysis techniques [29]. The singles and the electron-coincidence gamma data at -900V sample bias were collected simultaneously in one day. (b) "Singles" Doppler broadening spectrum collected at a -20 kV sample bias compared to the spectrum measured with the gamma-gamma coincidence condition. The HPGe signals were collected in coincidence with the annihilation gamma induced pulses from a NaI scintillation detector. The singles spectrum was collected using semi-analog nuclear electronics, whereas the coincidence spectrum was collected digitally. The coincidence condition reduces the high energy background by an order of magnitude and is consistent with other similar setups [39] due to the rejection of events in which both 511 keV gamma rays (one of them previously scattered) are detected at the same detector. The gamma-gamma coincidence data with -20kV on the sample were collected in three days.

### 3.3. Gamma- electron Coincidence

By extracting the time pick-off signal for the arrival of the annihilation gamma photon from the HPGe detector output, it is possible to correlate the ToF of the positron-induced secondary electrons, as well as the positron annihilation-induced Auger electrons, to the energy of the Doppler-shifted annihilation gamma photon. The coincident measurement of gamma energy and electron ToF results in a 2D correlation matrix and can yield such critical information as the Doppler broadening associated with the annihilation of positrons with specific core electronic levels. The results of the gamma-electron coincidence measurements are discussed below.

#### 3.3.1. Gamma-secondary electron coincidence

Figure 12(a) shows histogram of the measured time difference ($\Delta t$) between the detection of the secondary electrons emitted following the implantation of ~ 56 eV positrons and the detection of the concurrent gamma produced after the annihilation of positron. The spectrum was measured with the low-energy settings (maximum positron energy ~ 6 eV) and a sample bias of - 50 V with respect to the ToF tube. The histogram of measured time differences was constructed using the time difference between the digitally acquired signals from the HPGe detector and the MCP detector [29]. The top axis of the figure shows the kinetic energy corresponding to the measured time difference between the gamma and the electron signals obtained using the energy calibration described in the preceding section. By energy conservation, the maximum kinetic energy of the secondary electrons detected at the MCP should only be ~ 106 eV. However, there is significant intensity below the flight times of electrons (~ 555 ns) ejected with the maximum possible energy (106 eV) allowed by energy conservation. This anomaly is due to the association of the secondary electrons with the gamma emitted by the delayed annihilation of the spin-triplet state of positronium (o-Ps). The decay of o-Ps will be reflected in the spectrum as an exponentially decaying tail that extends even up to negative flight times that corresponds to the detection of secondary electrons prior to the detection of the delayed annihilation gamma. The exponential tail of the secondary electron spectrum, and the corresponding change in the slope of the tail, is emphasized by plotting the secondary electron intensity in the natural logarithmic scale.

The annihilation gamma-induced signal from the HPGe that was utilized in the construction of the ToF spectrum was also used to construct the energy spectrum of the Doppler-shifted annihilation gamma photons (Figure 12(b)). Hence, each electron ToF in Figure 12(a) is associated with a single annihilation gamma photon as shown in Figure 12(b). This correlation between the measured $\Delta t$ of the electron and the annihilation gamma energy can be expressed in terms of a 2D histogram with axes composed of the $\Delta t$ of the individual electrons and the energies of the corresponding annihilation photon. Figure 12(c) is a heat map of the two-dimensional correlation histogram, and Figure 12(d) expresses the two-dimensional histogram as a three-dimensional surface plot. A projection of 2D histogram along the time axis can be seen to be equal to the data shown in Figure 12(a); correspondingly, a projection along the gamma energy axis can be seen to equal to the data in Figure 12(b). It can be seen that most of the secondary electrons with energies between ~ 800 ns and ~ 555 ns (~ 50 eV to 106 eV) are associated with the 511 keV photo-peak and the associated Compton region; whereas $\Delta t$ less than 555 ns are associated solely with o-Ps decay, as evident by the small intensity in the photo-peak region. Hence, a summation along the $\Delta t$ bins from negative times to 555 ns will yield a gamma spectrum due entirely to o-Ps annihilations, including pick-off annihilations, particular to our spectrometer. Similarly, a summation along gamma bins corresponding only to the photo-peak will yield a secondary spectrum entirely devoid of the o-Ps decay tail [40].

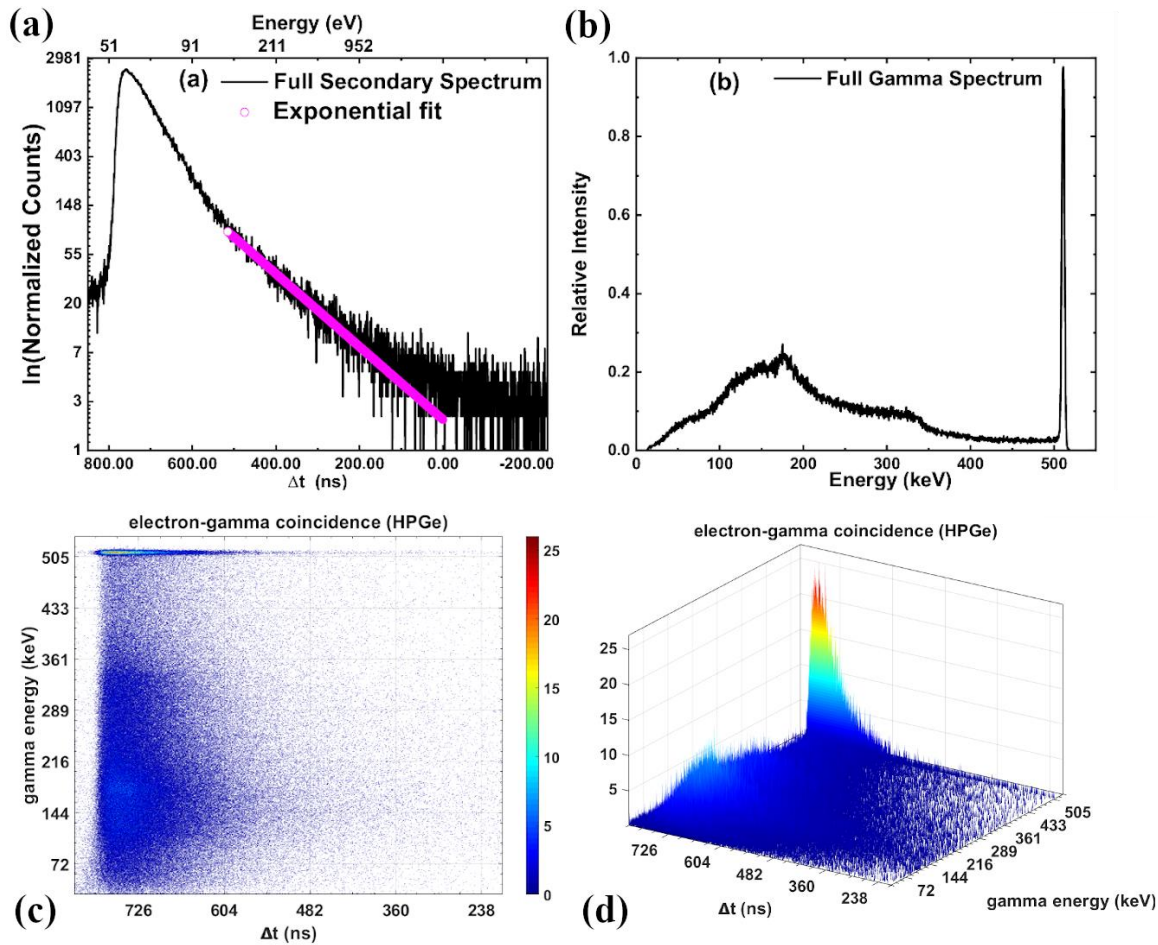

**Figure 12**. (a) Histogram of the measured time difference (Δ$t$) between the detection of the secondary electrons emitted following the implantation of ~ 56 eV positrons and the detection of the concurrent gamma produced after the annihilation of the positron. The spectrum was measured with the low-energy settings (positron energy < 6 eV) and a sample bias of -50 V with respect to the ToF tube in two days. The sample is bilayer graphene on polycrystalline Cu. The top axis of the figure shows the kinetic energy corresponding to the measured Δ$t$. Δ$t$ between 800 ns and 555 ns corresponds to the secondary electron peak (min. and max. electron energy) and the region < 555 ns signifies the correlation of an emitted electron with the delayed annihilation of o-Ps. The correlation of the secondary electrons with the delayed gamma photon emitted following the annihilation of o-Ps has been demonstrated by extracting the o-Ps lifetime using an exponential fit to the tail of the secondary spectrum (red circles). The fit gave the o-Ps decay rate including pick-off annihilations to be 7.44± 0.1 µsec$^{-1}$ (~134 ns). (b) Annihilation gamma spectrum generated using the digital pulse height analysis of the HPGe signals which were also used for the determination of the secondary spectrum shown in *(a)*. Each Δ$t$ in *(a)* is associated with a single annihilation gamma photon. (c) Heat map representation of the coincident measurement of gamma photons and electrons resulting from positrons incident on the sample. The gamma spectrum corresponding to Δt < 555 ns are due to the annihilation of o-Ps. By using this correlation, we can extract gamma spectrum originating entirely from o-Ps decay including pick-off annihilations. (d) The same data shown in Panel *(c)* but expressed as a three-dimensional surface plot and rotated to clarify the 511 keV gamma spectrum and secondary electron peak embedded within.

The three-dimensional surface plot presented here is very similar to the output of Age-Momentum correlation (AMOC) measurements, in which the positron lifetime is correlated with the Doppler-broadened annihilation gamma spectrum [41-42]. In the secondary electron-gamma coincidence measurements described, the "age", or positron lifetime, is the time between the detection of the annihilation gamma and the electron. When the sample is biased to high potentials such that the electrons are travelling at energies > 1 keV, the secondary electron-gamma coincidence measurement described here can, in principle, yield similar information as that obtained with beam-based AMOC measurements. We demonstrate the similarity with AMOC by fitting the exponentially decaying tail below measured time difference of 555 ns to obtain the o-Ps decay rate including pick-off annihilations to be $7.44 \pm 0.1$ μsec$^{-1}$ which compares well with the o-Ps decay rate of 7.0398 (29) μsec$^{-1}$ in vacuum [43]. With better counts in the tail and by using the gamma electron coincidence method, we will be able to gauge the percentage of pick-off annihilations in our system.

### 3.3.2. Gamma-Auger electron coincidence

The gamma-Auger electron coincidence spectrum, measured with low energy positrons (< 6 eV) on a sample of bilayer graphene on a Cu substrate, is shown as a 2D heat map in Figure 13(a). The 511 keV photo-peak can be identified when viewing the figure along the gamma axis. The presence of C-KVV and O-KVV peaks can also be discerned when viewing along the time axis. Figure 13(b) shows the heat map in an expanded scale to clearly indicate the presence of the embedded Auger peaks between 650 ns and 200 ns (C-KVV and O-KVV) along the region corresponding to the 511 keV photo-peak. The expanded 2D surface plot shows clearly that the Doppler broadening of the 511 keV photo-peak, for electron flight times between ~ 650 ns and 200 ns, is greater than the region corresponding to flight times > 650 ns. This broadening can be explained by considering the following. Electron flight times between 650 ns and 200 ns correspond to the high-energy core Auger peaks from carbon (C-KVV at 263 eV) and oxygen (O-KVV at 503 eV) atoms on the sample surface. The KVV Auger electrons are emitted following the Auger decay of the K-shell hole created by the annihilation of a positron with a K-shell electron in carbon or oxygen. The maximum Doppler broadening of the annihilation gamma photons is approximately proportional to the square root of the binding energy of the

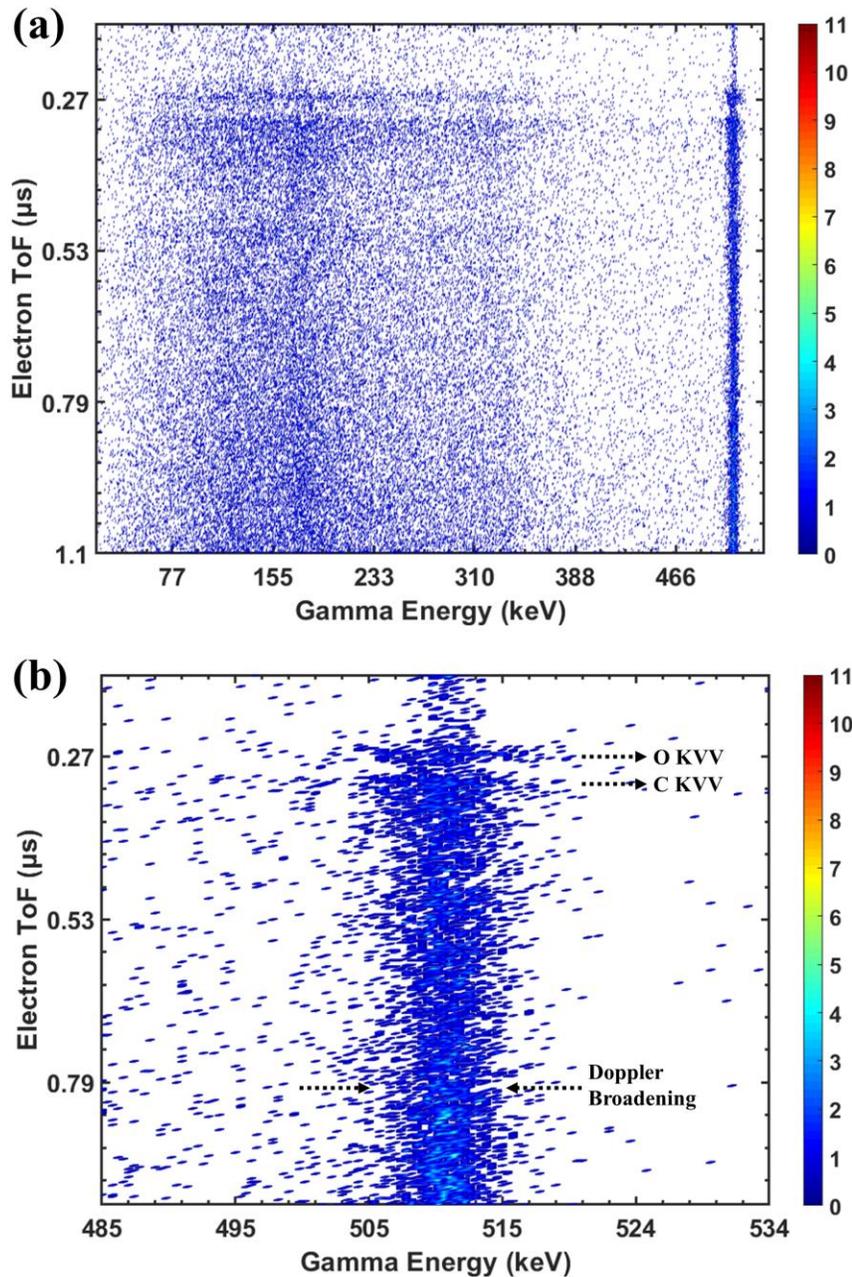

**Figure 13** (a) Heat map representation of the gamma-Auger electron coincidence spectrum, measured with low energy positrons (< 6 eV) on a sample of bilayer graphene on a Cu substrate. (b) The heat map shown in *(a)* but in an expanded scale. The Auger electron peaks embedded between 650 ns and 200 ns on the Doppler broadened 511 keV photo-peak can be seen in the expanded scale. The peak positions of the C KVV and O KVV Auger peaks are shown by dashed arrows for easy identification. The Doppler broadening of the 511 keV photo-peak (as indicated in the figure), for core Auger electron flight times between ~ 650 ns and 200 ns, is greater than the region corresponding to flight times > 650 ns. The gamma photons associated with Auger electrons represent core annihilations and are therefore Doppler-shifted by a larger amount than those gamma photons that are associated with valence or shallow core levels.

electron with which the positron annihilates; hence, the gamma photons associated with core Auger electrons will be emitted within a broader range of energies than the gamma photons associated with valence or shallow core annihilations. This concept forms the basis of Doppler broadening spectroscopy of open volume defects in materials [1-4].

The Doppler-broadened annihilation gamma peak associated with carbon K-shell annihilations alone can be extracted by taking a cut along the C-KVV Auger peak in the ToF spectrum. We define a cut, or a limited projection, as a one-dimensional histogram obtained by integrating the counts with respect to one variable of the original two-dimensional histogram over a defined range. In this case (data shown in Figure 14 (b)) we integrated with respect to the ToF variable over the range corresponding to C KVV Auger peak. We use the word "projection" to signify an integration over the full measured range of one variable. The projection of the 2D correlation matrix along the ToF axis yields the full PAES spectrum as shown in Figure 14(a). The region along which the coincidence spectrum was cut has been marked with a red background in Figure 14(a). The resulting annihilation gamma photo-peak is shown in Figure 14(b), along with the annihilation spectrum measured without any coincidence condition and at the same beam settings. The coincidence gamma spectrum obtained after taking the C-KVV cut was corrected for the presence of accidental coincidences by considering the individual singles count rate of the MCP and the HPGe detector [33]. Only the high energy region ($\geq 511$ keV) of the background-corrected photo-peak was used to produce the Doppler broadening spectrum shown. This was to avoid the lower-energy background contributions due to inefficient charge collection in the detector. A symmetric Doppler peak was constructed by taking the reflection of the high-energy region of the Doppler-broadened annihilation photo-peak about the point corresponding to 511 keV along the y-axis. The cut Doppler-broadened spectrum corresponds to positron annihilations only with electrons belonging to the K shell of carbon atoms on the top surface. On the other hand, the no-coincidence singles spectrum contains Doppler-shifted annihilation gamma photons from both valence and core annihilations. The singles spectrum is, however, dominated by valence annihilations because of the reduced overlap of the positron wavefunction with electron wavefunction belonging to the core electronic states [5-7, 44] and therefore should exhibit less broadening than the spectrum associated with the K-shell annihilations. Figure 14(b) shows that the Doppler-broadened spectrum from K-shell annihilations is significantly broader than the singles spectrum: the FWHM of the K-shell annihilation spectrum is 1.5 times the FWHM of the singles spectrum. Figure 14(b) demonstrates the capability of the gamma-electron coincidence spectrometer to measure the spectrum of the Doppler-shifted gamma photons originating from the annihilation of positrons with specific core levels whose subsequent decay may result in an Auger peak.

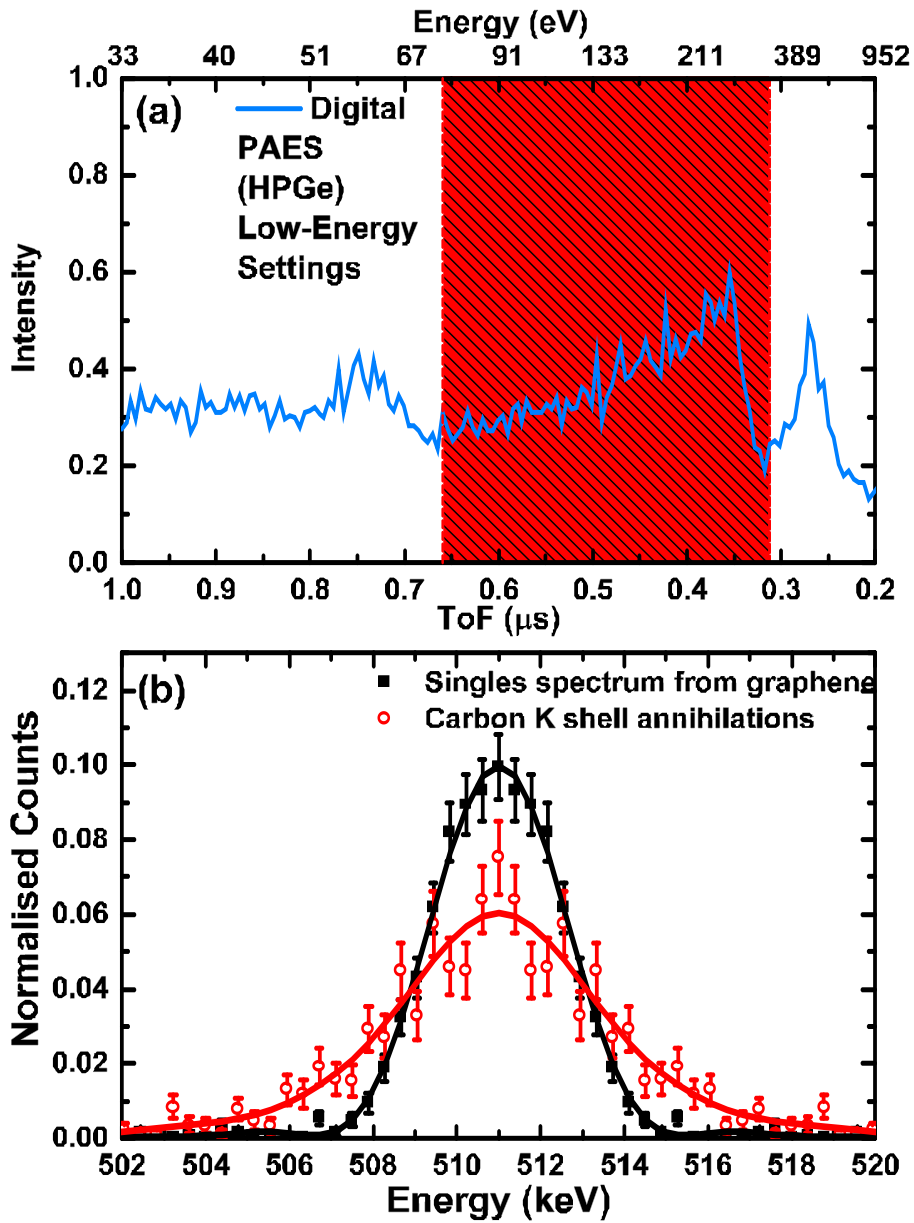

**Figure 14** (a) The full PAES time-of-flight spectrum obtained by taking a projection of the 2D correlation matrix along the ToF axis. The red shaded box denotes the region along which the "cut" is taken to obtain the Doppler-broadened annihilation gamma spectrum produced due to the annihilation of positrons exclusively with the K shell electrons of carbon. The PAES spectrum has been area normalized. (b) Comparison of the annihilation gamma spectrum associated with carbon K-shell annihilations with the singles gamma spectrum (collected using the same settings without any coincidence condition). The FWHM of the K-shell annihilation spectrum is 1.5 times the FWHM of the singles spectrum. The solid lines through the spectra are an aid to the eye generated by smoothing the experimental data using an FFT filter method to remove the high-frequency fluctuations caused by the limited statistics of the spectrum.

### 3.4. Energy spectra of secondary electrons measured using the multi-stop ToF spectrometer

By digitizing the acquisition of signals from MCP, it is possible to acquire all MCP signals associated with a single annihilation or positron event. Thus, the digitization of the electron ToF spectroscopy allowed us to convert our conventional "single- or first-electron" ToF spectrometer [11] into a multi-stop ToF spectrometer, in which all electrons emitted following positron-impact or annihilation are measured. The multi-stop capability of the digital ToF spectrometer is demonstrated in Figure 15. Figure 15(a) shows the energy spectrum of secondary electrons ejected from bilayer graphene on Cu when ~ 106 eV positrons are incident on the sample. The energy histogram has been constructed from the digitally measured time difference between the detection of the gamma ray by the HPGe detector and the electron by the MCP. While measuring the time difference between the gamma and the electron signals, only the "fastest" pulse from the MCP was considered (equivalent to the workings of an analog single-stop TAC with a window of 5 µs). We call result of this the single- or first-electron spectrum, and it has a shape typical of secondary electron spectra [35]. The spectrum has been normalized to the total number of events.

Figure 15(b) shows the spectra that result when two MCP pulses are detected in coincidence with a single annihilation event. The two-electron spectrum has been derived from the same data set as the single- or first-electron spectrum by selecting only those events in which two (and exactly two) MCP signals are detected within the coincidence window of 5 µs. We identify the two electrons as the first electron and the second electron and have plotted their corresponding spectra separately in Figure 15(b). The spectra of the two-electron events have been normalized to the area under the single- or first-electron spectrum in Figure 15(a) to give a sense of the probability of detecting a pair of electrons. This probability, in comparison to the detection of single electron events, has been shown recently to be a measure of the electron correlation strength of the material by Brandt, et al. [19]. In their experiment, the energy of the positron impact-induced two-electron events (*p,2e*) were measured using two hemispherical analysers. The major difference between the measurement in [19] and the data shown here is that the pair emission in the present spectra is in coincidence with the annihilation gamma, whereas the Brandt, et al. measured the electrons without the detection of the corresponding backscattered positron or the positron-induced annihilation gamma. In the present experiment, we detect the annihilation gamma in coincidence with the electrons, and therefore the detected electron pair is associated with a positron that either annihilated in the sample or was emitted as Ps. Correspondingly, this electron pair is not associated with a positron scattered out of the material, as the annihilation of backscattered positrons will occur outside of the HPGe detector's solid angle of detection.

The energy distribution of the first and the second electron is consistent with energy conservation. The energy distribution of first electron has significant intensity only below ~ 70 eV, whereas the energy distribution of the second electron has a high-energy cut-off at ~ 25 eV. This is consistent with the total energy available for two electrons with respect to the vacuum

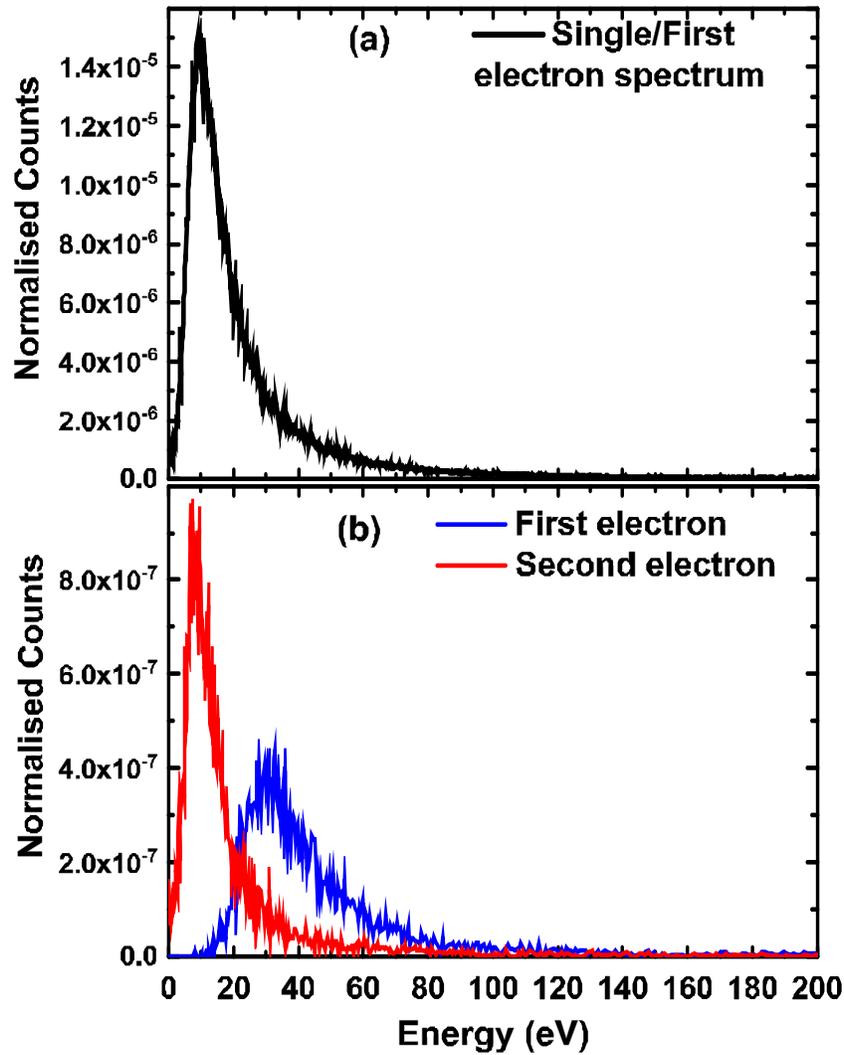

**Figure 15** (a) Energy spectrum of secondary electrons emitted following the implantation of ~ 106 eV positrons on bilayer graphene on a substrate of Cu. The 3 m ToF spectrometer was used to create a histogram of the time difference between the detection of the annihilation gamma by the HPGe detector and the detection of the electron by the MCP. The histogram was converted into electron energy spectrum using the energy calibration described earlier. Here, only the fastest MCP signal was considered for the digital construction of the time histogram, equivalent to the method employed in a single stop TAC; hence, the spectrum has been identified as resulting from the detection of a single/first electron. The spectrum has been normalized to have unit area. (b) Energy spectrum of the first electron and the second electron emitted following the implantation of 106 eV positrons. The spectra have been divided by the total counts under the single/first electron spectrum. The energy spectra were constructed by considering only those events in which two MCP pulses occurred within 5 µs of the detection of the annihilation gamma photon. The two-electron spectra are consistent with energy conservation and demonstrates the capability of the new multi-stop ToF spectrometer. The measurement of multielectron secondary spectra took five days to complete.

level (96 eV, approximately the maximum kinetic energy of the primary positron minus two work functions). The energy distribution of the second electron extends down to 0 eV, whereas the low-energy region of the first electron cuts off at ~ 10 eV. For two MCP pulses to be distinguished, the time difference between the two should be greater than half the total width of the MCP pulse. This requirement introduces the difference in the low energy cut-off seen in the 100 V spectrum. However, for lower energy electrons (i.e. lower sample biases), the low energy cut-off shift arising from the finite width of the MCP pulses will become negligible. This is due to the small energy differences of the electrons resulting in a time separation of the pulses that is larger than half the width of an individual MCP pulse [29].

### 3.5. *Applications of the advanced positron beam*

The multifunctional nature of the advanced positron beam allows it to characterize the topmost atomic layer of surfaces as well as the sub-surface region, hidden interfaces, and inner surfaces of porous materials. The ability of Doppler broadening spectroscopy to selectively characterize the chemical nature of the topmost layer of the external surfaces was demonstrated recently using the advanced positron beam [45-46]. These investigations provided the proof-of-principle that chemical characterization using Doppler broadening spectroscopy provides the same level of surface selectivity as that obtained with PAES [13]. Based on these results, it was suggested that Doppler broadening spectroscopy can be used for the characterization of the top-most atomic layers of a porous materials' inner surfaces with feature sizes in the nano-meter scale. Since the energy of the positrons can be varied from a few eV to 20 keV, the advanced positron beam can also investigate defects at the surfaces and interfaces of thin films that are only a few nano-meters thick [47].

Since the energy of the positrons that are incident on the sample can be reduced to a couple of eV, it is possible to obtain positron annihilation-induced Auger electron spectra without a probe-induced secondary electron background [15]. The absence of a low-energy secondary background has recently resulted in the direct measurement of low-energy Auger processes (< 15 eV) which are unobservable by other electron- or photon-induced electron emission spectroscopies [13, 37-38]. This ability of PAES has also led to the discovery of exotic low-energy electron emission processes such as Auger-mediated positron sticking (AMPS) [13], which is expected to provide direct information on the surface-projected density of states without any contribution from the bulk density of states [48]. We expect that the advanced positron beam may be able to provide insight into other process that are accompanied by low-energy electron emission, including positron annihilation-induced ion desorption [18] and Ps formation through the Auger neutralization of a positron at the surface.

The gamma-Auger electron coincidence spectroscopy technique described in Section 3.3.2 can extract the contribution of the momenta of specific core electronic levels to the total Doppler broadening of the annihilation gamma spectrum. This provides a method to experimentally benchmark the ability of existing ab-initio techniques that aims to accurately predict the characteristics of a surface-trapped positron [44]. The gamma spectra associated with Auger electron emission have zero contribution from Ps formation; whereas the gamma photons that are associated with the exponentially decaying tail of the secondary electron spectrum, as shown in Figure 12(a), are entirely due to Ps formation. This fact, along with gamma-electron coincidence method, can potentially be used to extract gamma spectra with 0% and 100% Ps contribution to accurately measure the Ps formation fraction of porous materials.

Multiple electron emission following the Auger decay of a core hole forms the foundation for explaining the extra intensity found in the inelastic tail of Auger peaks originating from shallow core levels [20-24]. The intensity and energy distribution of the multiple Auger electrons from different samples can provide an estimate of the strength and role of correlation of core holes with valence holes and valence holes with other valence holes in the photon-, electron- or positron-induced electron emission process. The positron beam presented here is equipped with a digitized multi-stop ToF spectrometer and can selectively extract low-energy electrons associated only with specific higher-energy Auger peak. This feature will potentially provide an accurate description of the multi-electron Auger decay process; the details of which will be discussed in an upcoming publication [49].

## 4. CONCLUSIONS

We have developed an advanced multi-functional, variable-energy positron beam system that can measure the energies of multiple "positron-induced" electrons in coincidence with the Doppler-shifted gamma photon resulting from the annihilation of the correlated positron. The measurements were made possible by the 3 m ToF path along with digital data acquisition and analysis methods. These innovations have led to the first digital measurement of positron annihilation-induced Auger electrons generated using coincident signals from a high-purity Ge detector (for annihilation gamma rays) and a micro-channel plate (for positron-induced electrons). The coincident measurements were used to develop a two-dimensional spectrum that provided the energy of Doppler-broadened annihilation gamma photons and the energy of either the: (i) positron annihilation-induced Auger electrons or (ii) positron impact induced secondary electrons that were emitted concurrently. The analysis of the two-dimensional gamma energy – Auger electron ToF coincidence spectrum resulted in the spectrum of Doppler-shifted gamma photons due to the annihilation of positrons with 1s electrons of carbon. The digitization of the ToF spectrometer led to the development of a multi-stop ToF spectrometer and was utilized to obtain the energy distribution of multiple secondary electrons ejected from a bilayer graphene surface as a result of the positron impact. These advancements have resulted in a positron beam that can selectively probe the top-most atomic layers of a material and investigate the sub-surface regions and the inner surfaces of nanoporous materials. These capabilities will provide information concerning the surface-projected electronic structure, the electron correlation strength, and the chemical characteristics and defect architecture of novel materials.

**ACKNOWLEDGMENTS**

The advanced positron beam system at UT Arlington was developed using the NSF major research instrumentation grant DMR-1338130. The research work on two dimensional materials was funded by NSF grant DMR-1508719. We gratefully acknowledge the support by Welch Foundation (grant No. Y-1968-20180324) for the development of coincidence methods described in the manuscript.